%% file: hierarchical_beamforming_arxiv.tex
\pdfoutput=1

\documentclass{article}

\input{config}
\usepackage{booktabs} 
\begin{document}
\title{Hierarchical Beamforming: Resource Allocation, Fairness and Flow Level Performance
\author{Julien Floquet ($\dag,\times$), Richard Combes ($\dag$) and Zwi Altman ($\times$)}
\thanks{$\dag$: Centrale-Supelec and L2S, Gif-sur-Yvette, France}
\thanks{$\times$: Orange Labs, Paris, France}} 
\renewcommand\footnotemark{}
\renewcommand\footnoterule{}
\maketitle
\begin{abstract}
We consider hierarchical beamforming in wireless networks. For a given population of flows, we propose computationally efficient algorithms for fair rate allocation including proportional fairness and max-min fairness. We next propose closed-form formulas for flow level performance, for both elastic (with either proportional fairness and max-min fairness) and streaming traffic. We further assess the performance of hierarchical beamforming using numerical experiments. Since the proposed solutions have low complexity compared to conventional beamforming, our work suggests that hierarchical beamforming is a promising candidate for the implementation of beamforming in future cellular networks.
\end{abstract}
\input{hierarchical_beamforming_body}
\end{document}

%% file: config.tex
\makeatletter
\DeclareRobustCommand*\cal{\@fontswitch\relax\mathcal}
\makeatother
\usepackage[utf8]{inputenc}
\usepackage{dsfont}
\usepackage{amsmath}
\usepackage{amsthm}
\usepackage{enumerate}
\usepackage{amsmath}
\usepackage{amssymb}
\usepackage[ruled,vlined]{algorithm2e}
\usepackage{graphicx}
\usepackage{subfig}
\usepackage{epstopdf}
\usepackage{multicol}
\usepackage{longtable}
\usepackage{amsfonts}
\usepackage{textcomp}
\usepackage[nolist]{acronym}
\usepackage{color}
\usepackage{manfnt} 
\usepackage{verbatim} 
\usepackage{bm}

\newcommand{\indic}{ {\bf 1}}

\newcommand{\floor}[1]{  \lfloor #1 \rfloor}

\newcommand{\als}[1]{ \begin{align*} #1  \end{align*}}

\def\conv{{\bf conv}}
\def\deg{d}

\def\bp{\noindent {\it Proof.}\ }
\def\ep{\hfill $\Box$} 
\def\indic{{\bf 1}} 
\def\bi{\begin{itemize}}
\def\ei{\end{itemize}}

\def\CC{{\mathbb C}}

\def\EE{{\mathbb E}}

\def\NN{{\mathbb N}}

\def\PP{{\mathbb P}}

\def\RR{{\mathbb R}}

\def\cA{{\cal A}}

\def\cD{{\cal D}}

\def\cG{{\cal G}}

\def\cK{{\cal K}}

\def\cM{{\cal M}}
\def\cN{{\cal N}}
\def\cO{{\cal O}}

\def\cQ{{\cal Q}}

\def\cX{{\cal X}}
 
\def\cZ{{\cal Z}}
\def\bblambda{{\boldsymbol \lambda}}
\def\bbrho{{\boldsymbol \rho}}
\def\bbphi{{\boldsymbol \phi}}
\def\bbtheta{{\boldsymbol \theta}}
\def\bbkappa{{\boldsymbol \kappa}}
\def\bbdelta{{\boldsymbol \delta}}
\def\bbgamma{{\boldsymbol \gamma}}

\def\bbe{{\bf e}}

\def\bbm{{\bf m}}
\def\bbn{{\bf n}}

\def\bbp{{\bf p}}

\def\bbr{{\bf r}}
\def\bbs{{\bf s}}

\def\bbx{{\bf x}}

\def\bbz{{\bf z}}

\def\bbM{{\bf M}}
\def\bbN{{\bf N}}

\def\bbX{{\bf X}}

\newtheorem{defin}{Definition}

\newtheorem{corollary}{Corollary}
\newtheorem{theo}{Theorem}
\newtheorem{prop}{Proposition}

\newtheorem{fact}{Fact}
\begin{acronym}
    \acro{OFDMA}{Orthogonal Frequency-Division Multiple Access}
    \acro{NGMN} {Next Generation Mobile Network}
    \acro{WiMAX}{Worldwide Interoperability for Microwave Access}
    \acro{LTE}{Long Term Evolution}
    \acro{RAN}{Radio Access Networks}
    \acro{QoS}{Quality of Service}
    \acro{ICIC}{Inter-Cell Interference Coordination}
    \acro{LB}{Load Balancing}
    \acro{PC}{Power Control}
    \acro{FFR}{Fractional Frequency Reuse}
    \acro{FL}{Fractional Load}
    \acro{BCR}{Block Call Rate}
    \acro{eNB}{enhanced eNode B}
    \acro{CDMA}{Code Division Multiple Access}
    \acro{GSM}{Global System for Mobile Communications}
    \acro{BS}{Base Station}
    \acro{PRB}{Physical Resource Block}
    \acro{NG}{Next Generation}
    \acro{B3G}{Beyond 3G}
    \acro{SON}{Self-organizing networks}
    \acro{3GPP}{3rd Generation Partnership Project}
    \acro{UE}{User Equipment}
    \acro{RRM}{Radio Resource Management}
    \acro{KPIs}{Key Performance Indicators}
    \acro{SINR}{Signal to Interference plus Noise Ratio}
    \acro{HSDPA}{High Speed Downlink Packet Access}
    \acro{RR}{Round Robin}
    \acro{TDMA}{Time Division Multiple Access}
    \acro{MTP}{Max Throughput}
    \acro{PF}{Proportional Fair}
    \acro{MMF}{Max-Min Fair}
    \acro{MAB}{Multi-Armed-Bandit}
    \acro{RB}{Restless Bandit}
    \acro{ODE}{Ordinary Differential Equation}
    \acro{i.i.d}{independent and identically distributed}
    \acro{MIMO}{Multiple Input Multiple Output}
    \acro{c.d.f}{cumulative distribution function}
    \acro{a.s}{almost surely}
    \acro{a.e}{almost everywhere}
    \acro{p.d.f}{probability density function}
    \acro{PPRBS}{Per Physical Resource Block Scheduler}
    \acro{KPI}{Key Performance Indicator}
    \acro{SIC}{Successive Interference Cancellation}
    \acro{FTP}{File Transfer Protocol}
    \acro{FTT}{File Transfer Time}
    \acro{LoS}{Line of Sight}
	\acro{MDP}{Markov Decision Process}
	\acro{CTMDP}{Continuous Time Markov Decision Process}
	\acro{DTMDP}{Discrete Time Markov Decision Process}
	\acro{POMDP}{Partially Observable Markov Decision Process}
	\acro{RS}{Reference Signal}
	\acro{BS}{Base Station}
	\acro{FCFS}{First Come First Served}
	\acro{PS}{Processor Sharing}
	\acro{HetNet}{Heterogeneous Network}
	\acro{SDP}{Semi-Definite Programming}
	\acro{iff}{if and only if}
	\acro{SON}{Self-Organizing Network}
	\acro{ODE}{Ordinary Differential Equation}
	\acro{w.r.t.}{with regard to}
	\acro{LHP}{Left Half Plane}
	\acro{SA}{Stochastic Approximation}
	\acro{OMC}{Operation and Maintenance Center}
	\acro{OSS}{Operation and Support System}
	\acro{NMS}{Network Management System}
	\acro{eNB}{\ac{BS}}
	\acro{3GPP}{3rd Generation Partnership Project}
	\acro{LMI}{Linear Matrix Inequality}
	\acrodefplural{LMI}[LMIs]{Linear Matrix Inequalities}
\acro{eICIC}{enhanced Inter Cell Interference Coordination}
\acro{CRE}{Cell Range Extension}
\acro{CIO}{Cell Individual Offset}
\acro{ABS}{Almost Blank Subframe}
\acro{ABSr}{ABS ratio}
\acro{ABSrO}{ABS ratio optimization}
\acro{CET}{Cell-Edge Throughput}
\acro{MUT}{Mean User Throughput}
\acro{CDF}{Cumulative Distribution Function}
\acro{PC}{Power Consumption}
\acro{EC}{Energy Consumption}
\acro{ViS}{Virtual Sectorization}
\acro{VSC}{Virtual Small Cell}
\acro{VeS}{Vertical Sectorization}
\acro{AAS}{Active Antenna Systems}
\acro{RSRP}{Reference Signal Received Power}
\acro{PEC}{perfect electrical conductor}
\acro{TR}{Time Reversal}
\acro{RxMF}{received matched filter}
\acro{TDD}{Time Division Duplex}
\acro{FDD}{Frequency Division Duplex}
\acro{MU}{Multi-User}
\acro{MTC}{Machine Type Communication}
\acro{AoA}{Angle of Arrival}
\acro{CQI}{Channel Quality Indicator}
\acro{CSI}{Channel State Information}
\acro{DL}{downlink}
\acro{UL}{uplink}
\acro{MCS}{Modulation and Coding Scheme}
\acro{FPC}{Fractional Power Control}
\acro{FD}{Full Dimension}
\acro{DFT}{Discrete Fourier Transform}
\acro{LSAS}{Large Scale Antenna System}
\end{acronym}

%% file: hierarchical_beamforming_body.tex
\section{Introduction}\label{sec:intro}
Massive \ac{MIMO} technology is considered as one of the main pillars of 5G \ac{RAN}, providing means to considerably increase
spectral and energy efficiency \cite{osseiran2014scenarios}. In the coming years, large scale antenna systems having hundreds of radiating elements are expected to be deployed.

	Large scale antenna systems with highly focused beams raise the problem of control channels that, unless being precoded (beamformed), will not match coverage of data channels. In Release 10 of \ac{3GPP}, the concept of precoded Channel State Information Reference Signal (CSI-RS) has been introduced \cite{LTE_A_2012}. Release 13 has introduced the \ac{FD}-\ac{MIMO} with 3D beamforming supported by 2D antenna arrays, with typically dual-polarized $8 \times 8$ antenna arrays \cite{Full_Dimension_MIMO}. With the introduction of massive \ac{MIMO} in 5G networks, the concept of beam switching or beam sweeping has been proposed. The beams from a given grid of beams are transmitted (in the downlink) or received (in the uplink) in a time interval and in a predetermined way. Beam sweeping can be used in both \ac{TDD} and \ac{FDD} for control channels. Data is transmitted using the best beam reported by the mobile user in \ac{FDD}. It is noted that grid of beams can be useful in \ac{TDD} when poor channel or \ac{SINR} condition occurs.

Recently, the concept of hierarchical or multilevel codebook of beams (or codebook based beamforming) has been proposed. The term multi-resolution has been used as well and refers to the same concept. One of the motivations behind the idea of hierarchical codebooks is to reduce the signaling overhead of common channels while maximizing the beamforming gain. 

Several contributions on hierarchical beamforming are briefly described presently. The first example is a millimeter wave backhaul serving urban pico-cells \cite{hur2013Nokia_MultilevelBF_IEEE_TC}. In this work, a hierarchical codebook of beams is used to efficiently align a pair of receive - transmit beams. 
Hierarchical beamforming has also been studied for the \ac{RAN}. In \cite{alkhateeb2014multiresolutionCodebook} the authors develop a hierarchical codebook for the training beamforming vectors for millimeter wave cellular systems. The sparse nature of the microwave channel is exploited in order to develop a hybrid analog/digital low complexity precoding algorithm. 
\cite{xiao2017codebook} introduces a generalized detection probability metric for comparing the efficiency of codebooks. An optimization procedure that optimizes this metric by flattening the beam patterns is proposed.
The hierarchical codebook of beams can be designed to fit coverage needs of a cell allowing to further reduce the number of beams in the codebook, as described in \cite{tall2015multilevel}. Furthermore it is shown that the geometry and propagation parameters of the cell should be taken into account when designing the structure of the codebook. 
For example, in typical dense urban cells, a 3D hierarchical codebook can be used while in sub-urban environment, a 2D horizontal hierarchical codebook of beams is more appropriate. More references on hierarchical beamforming can be found in \cite{Multires_hierarchical_ref} which also describes how to design the hierarchical codebook using the \ac{DFT} matrix.	 

Scheduling algorithms and flow-level performance of massive \ac{MIMO} taking into account both the physical layer and the dynamic nature of the traffic is a challenging and important problem that has received close to no attention in the literature~\cite{khlass2014flow}. 
Furthermore, the design of efficient \ac{MU}-\ac{MIMO} scheduling algorithms for massive MIMO systems having a hierarchical structure is relatively untouched.
In this article we first derive an efficient scheduling algorithm to compute the $\alpha$-fair allocation, in time $\cO(|V| + K)$ where $|V|$ is the number of beams and $K$ the number of flows. This ensures that fair rate sharing is relatively easy to implement for hierarchical beamforming, while this is not the case for all MIMO systems. We then consider flow-level dynamics where flows arrive and depart dynamically. For elastic traffic with Proportional Fair allocation, we derive the expected flow throughput in closed form, and prove that the system performance is insensitive (i.e. it depends only on the expected flow size and the arrival rate). For streaming traffic, we show that there exists algorithms to compute the blocking probability in almost linear time $\cO\left( |V| \ln |V|  \right)$ so that the system is tractable. We conclude with some illustrative numerical experiments. 

The remainder of this article is organized as follows. In Section~\ref{sec:model} we describe the general system model. In Section~\ref{sec:fair_rate_allocation} we provide an efficient algorithm to compute the $\alpha$-fair allocation for a fixed population of flows. In Section~\ref{sec:elastic} we consider hierarchical beamforming with flow-level dynamics and elastic traffic. In Sections~\ref{sec:elastic_pf} and~\ref{sec:elastic_mt}, we derive the flow-level performance of the PF and MT allocation (respectively) for elastic traffic. In Section~\ref{sec:streaming} we turn to streaming traffic and give an efficient algorithm to compute the blocking rates in almost linear time. Finally we provide some numerical experiments in Section~\ref{sec:numerical} and Section~\ref{sec:conclusion} concludes the paper.

\section{The model}\label{sec:model}
\subsection{Hierarchical codebook}
We consider a cell which is a region in the plane $\cG \subset \RR^2$.  We consider a set $V$ of fixed beams which can be used for transmitting data to flows. To each beam $v \in V = \{1,\dots,|V|\}$ we associate $\cG_v \subset \cG$ which is the region covered by beam $v$, namely one can send data to a flow located in $\cG_v$ using beam $v$. We assume that there exists a directed graph $G = (V,E)$ where the set of vertices $V$ is the set of beams. Graph $G$ is assumed to be a directed tree. For ease of presentation, we assume that the nodes $V$ are sorted by increasing depth, so that $1$ is the root of $G$, and $(v,v') \in E$ implies that $v < v'$. We call the set of beams, the covered regions and the corresponding graph a codebook. For beam $v \in V$ and $x \in \cG$, we denote by $g_{v}(x)$ the signal power received by a flow located at $x$ using beam $v$. It is noted that in practice, if $x \notin \cG_v$, the value of $g_v(x)$ is negligible, but the received signal power is not strictly zero.

\begin{figure}[h!]
\begin{center}
\leavevmode
\subfloat[Beam Depth = 0]{%
\label{fig-uniweria-magma}
\includegraphics[width=0.45\columnwidth,trim = 1cm 1cm 1cm 1cm, clip]{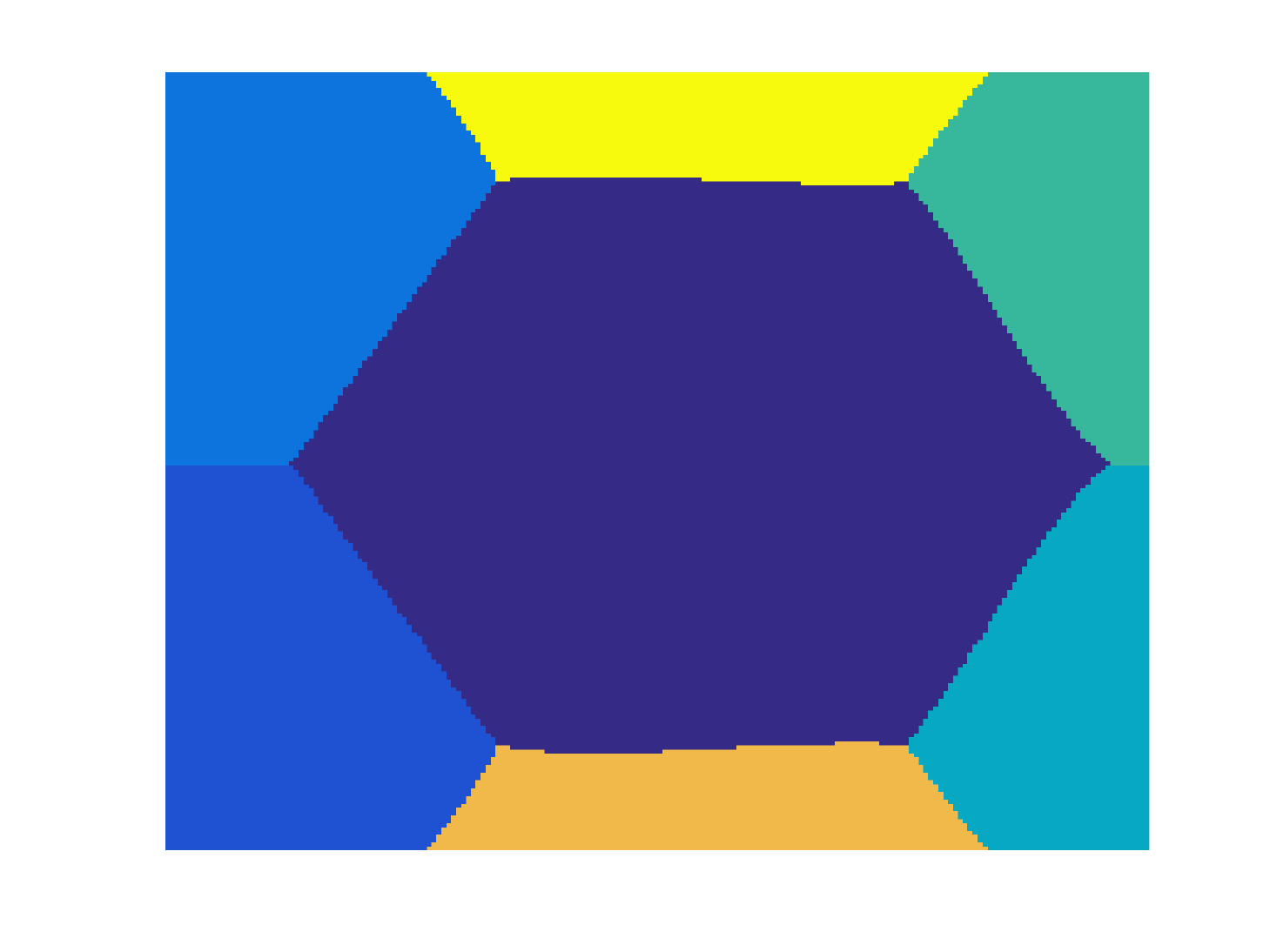}}
\subfloat[Beam Depth = 1]{%
\includegraphics[width=0.45\columnwidth,trim = 1cm 1cm 1cm 1cm, clip]{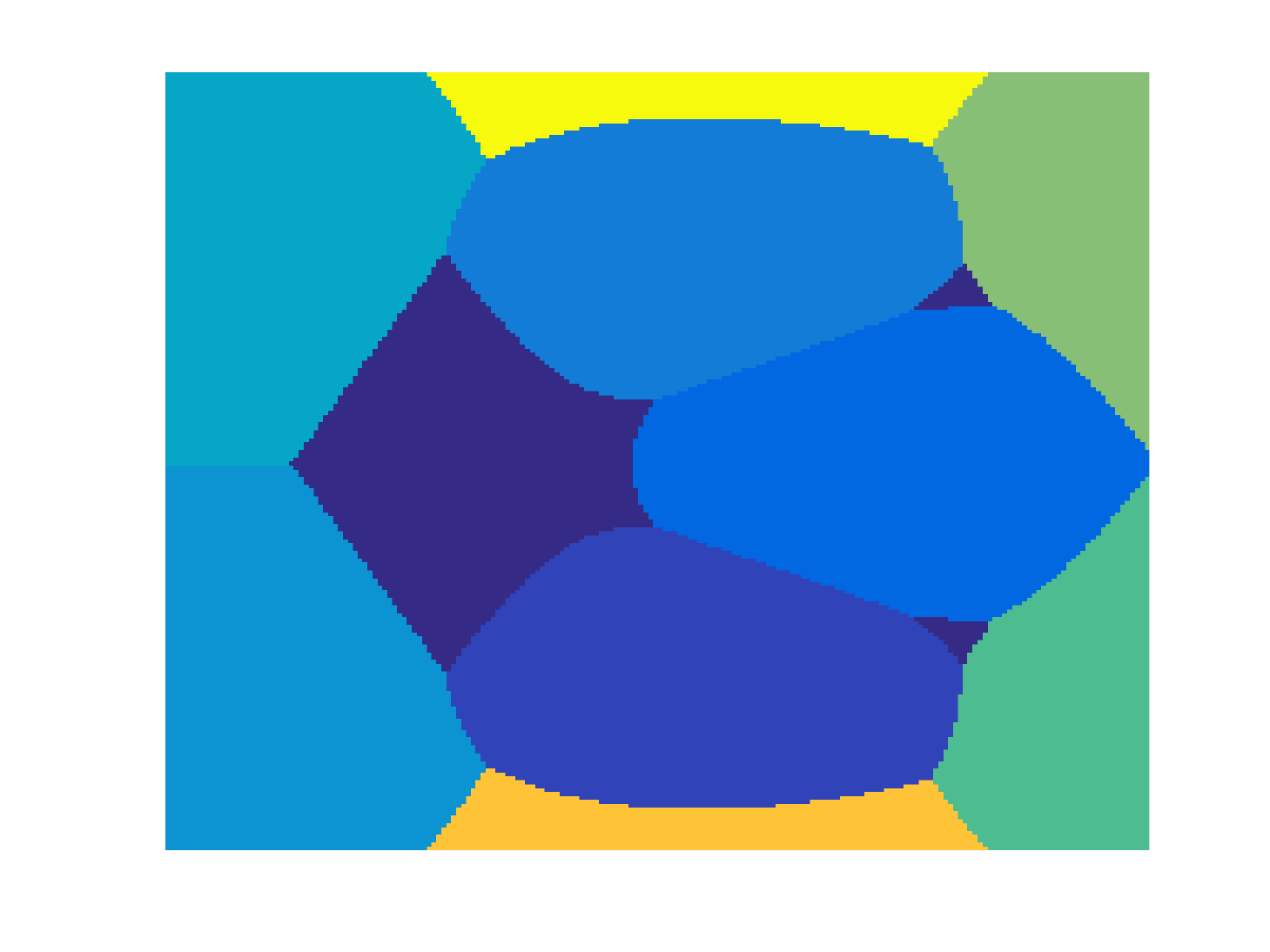}}
\\
\subfloat[Beam Depth = 2]{%
\includegraphics[width=0.45\columnwidth,trim = 1cm 1cm 1cm 1cm, clip]{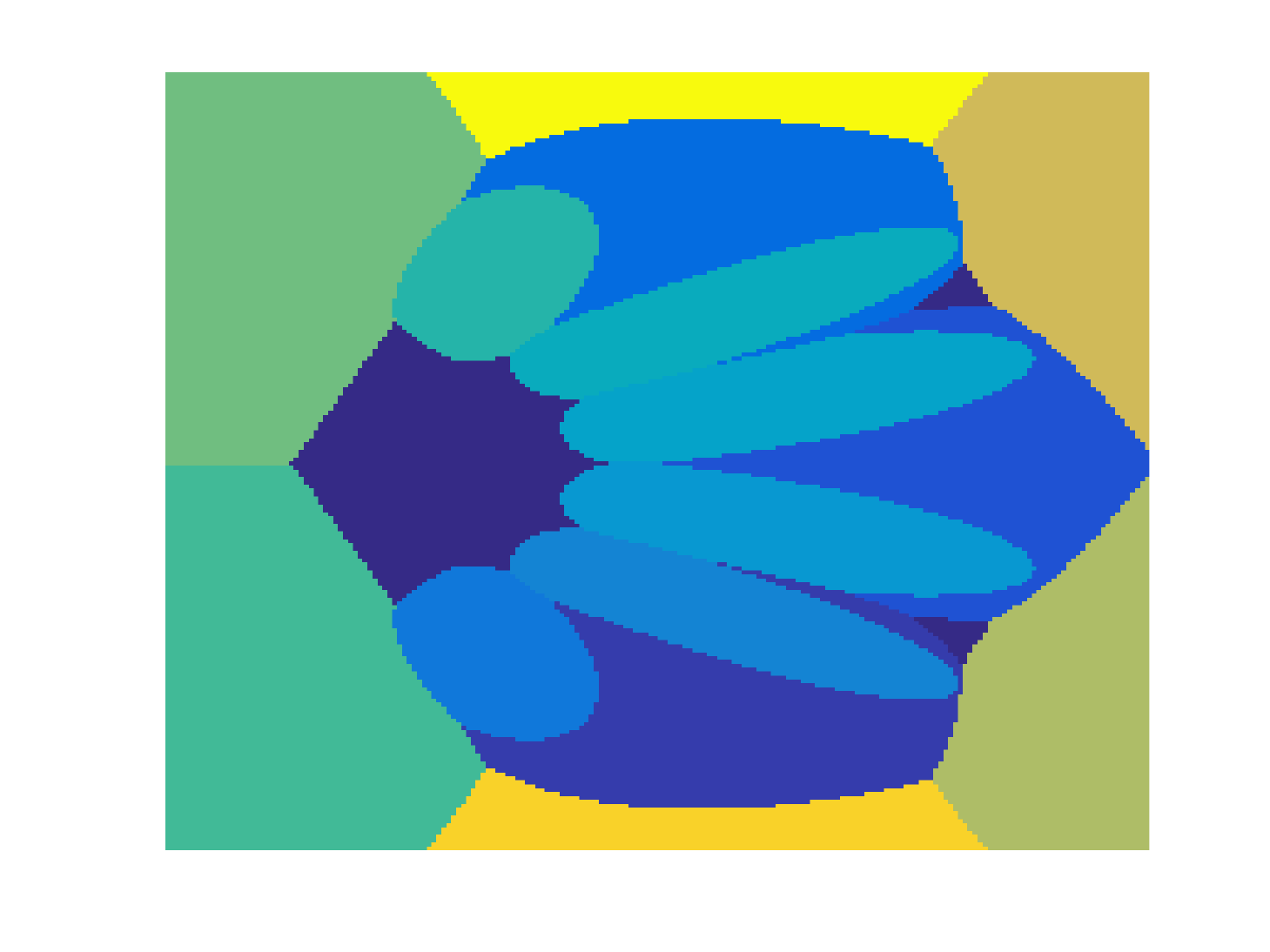}}
\subfloat[Codebook Graph]{\includegraphics[width=0.43\columnwidth]{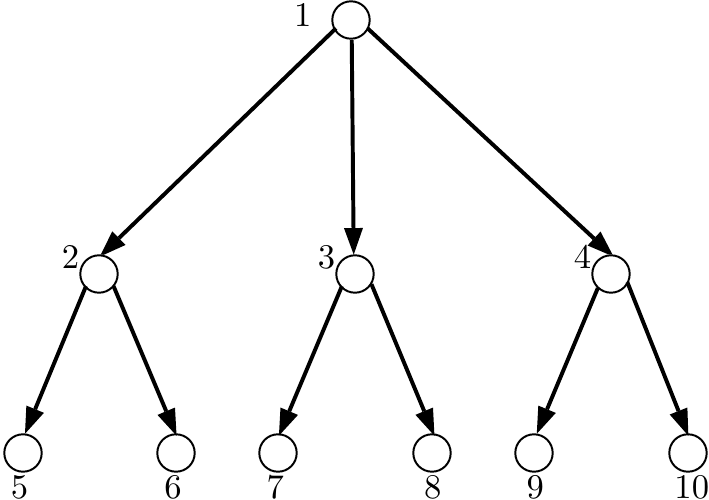}}
\caption{Hierarchical codebook example}
\label{fig:map}
\end{center}
\end{figure}

\begin{defin}\label{def:codebook} A codebook is hierarchical if it verifies the following three properties:
\bi
\item[(i)] If $(v,v') \in E$ then $\cG_{v'} \subset \cG_{v}$
\item[(ii)] If $(v,v') \in E$ and $(v,v'') \in E$ then $\cG_{v'} \cap \cG_{v''} = \emptyset$.
\item[(iii)] If $(v,v') \in E$ and $x \in \cG_{v'}$, then $g_{v'}(x) \ge g_{v}(x)$.
\ei
\end{defin}
An example of a hierarchical codebook with $|V| = 10$ beams is presented in Figure~\ref{fig:map}. Both the covered regions and the graph $G$ are depicted. 
As seen from the three above properties and the figure, the regions covered by beams form a hierarchical partition of space. Namely, the children of a beam $v$ cover regions covered by $v$ (property (i)), and the regions covered by distinct children do not overlap (property (ii)). Furthermore if one goes down in the tree, beams become more focused since they cover smaller regions, and the received signal power increases (property (iii)). We also see that hierarchical codebooks are an improvement over a grid-of-beams approach (where one only uses beams of maximal depth): the beams with the highest depth are small, very focused and do not interfere with each other (i.e. the interference is negligible), however any location not covered by those beams is covered by beams of lower depth. Beams of high depth allow for large data rates and beams of low depth ensure coverage to all flows in the cell. Therefore, while the grid-of-beams approach inevitably creates coverage degradation in between beams, hierarchical codebooks are immune to this problem.

\subsection{Notations used}
 We say that two beams $v$ and $v'$ do not interfere if and only if the regions they cover do not overlap, that is $\cG_v \cap \cG_{v'} = \emptyset$. We use the following notations:
\bi
\item $\cA(v)$ the set of ancestors of $v$, and $\bar{\cA}(v) = \cA(v) \cup \{v\}$.
\item $\cD(v)$ the set of descendants of $v$, and $\bar{\cD}(v) = \cD(v) \cup \{v\}$.
\item $\deg(v)$ the degree of $v \in V$ i.e. the number of its children
\item $\deg(G) = \max_{v \in V} \deg(v)$ the maximal degree of the graph $G$.
\item $h(G)$ the height of $G$ which is the largest distance between the root and any vertice $v \in V$.
\ei
The definition of the main graph-theoretic notions for trees are recalled in the appendix. We recall a standard fact about trees, which is that the number of edges equals the number of nodes minus one, $|E| = |V|-1$. Unless stated otherwise, we denote vectors by bold letters (e.g. $\bbx$), scalars by non-bold letters (e.g. $x$), and $x_k$ denotes the $k$-th component of $\bbx$. Deterministic quantities are denoted by lower-case letters (e.g. $x$) and random quantities are denoted by upper case letters (e.g. $X$). For instance $\bbX$ is a random vector, and $X_k$ is the $k$-th component of $\bbX$. Finally sets are denoted by upper case calligraphic letters (e.g. $\cX$). For $k$ an integer, we denote by $[k]$ the set $\{1,\dots,k\}$. We denote by $\ln$ the natural logarithm.

\section{Algorithms for fair rate allocation}\label{sec:fair_rate_allocation}

In this section we consider a fixed set $\cK$ of flows, where $x(k) \in \cG$, denotes the location of flow $k \in \cK$. We study fair rate allocation strategies. We first determine how flows should be associated to beams, and which sets of beams may be activated simultaneously. We then formulate fair rate allocation as a convex optimization problem and provide a problem-specific efficient algorithm to solve this problem.

\subsection{Flow association}

We assume that each flow $k \in \cK$ is associated to a beam $v_k \in V$ which maximizes the signal strength, so that 
\begin{equation*}
	v_k \in \arg\max_{v \in V} g_v(x(k)).
\end{equation*}
	Since the codebook is hierarchical, by property (iii), one should associate $k$ to the beam of largest depth which covers location $x(k)$. It is also noted that, from the hierarchy of the codebook, beam allocation to a flow given its location can be performed using a simple and efficient algorithm (Algorithm~\ref{algo:association}) described below. 
\begin{algorithm}
 \KwData{Tree $G$ sorted by decreasing height, point $x$, regions $\cG_v$ for $v \in V$}
 $v \leftarrow 1$;\\
 \While{$\exists v': (v,v') \in E, x \in \cG_{v'}$}{$v \gets v'$;}
 \KwResult{Flow at $x$ associated to beam $v$}
 \caption{Beam association algorithm}
 \label{algo:association}
\end{algorithm}
This algorithm runs in at most $\cO(d(G) h(G))$ iterations. For instance, if $G$ is a regular tree its height is $h(G) = \cO(\ln |V|)$, so that the running time is $\cO(d(G) \ln |V|)$. This is an advantage of hierarchical codebooks over the grid-of-beams approach, where flow association usually requires scanning through all the beams, which requires ${\cO}(|V|)$ time. Hence hierarchy allows a considerable improvement from linear to logarithmic complexity. This is critical for real-world implementation of large codebooks, especially since flow association must be updated periodically due to arrival, departure and mobility of flows.

When flow $k$ is served by beam $v_k$, it may receive data at data rate:
$$
r_k = W \log_2\left( 1 + {g_{v_k}(x_k) \over N_0^2}\right),
$$
where $W$ is the bandwidth and $N_0^2$ is the thermal noise power. We denote by $\cK(v) = \{k \in \cK: v_k = v\}$ the set of flows associated with beam $v$, and $n_v = |\cK(v)|$ the number of flows associated with beam $v$.

	While we do not take into account several cells in this model, we assume that, in a more realistic setting, if all neighboring cells use hierarchical beamforming as well, the resulting inter-cell interference should be negligible since when beamforming with narrow beams is used, two flows would see interference only if they are both physically close to each other and receive data at the same time. Indeed this situation should only happen infrequently. 

\subsection{Beam activation strategies}

\begin{figure}[!h]
\begin{center}
\includegraphics[width=0.75\columnwidth]{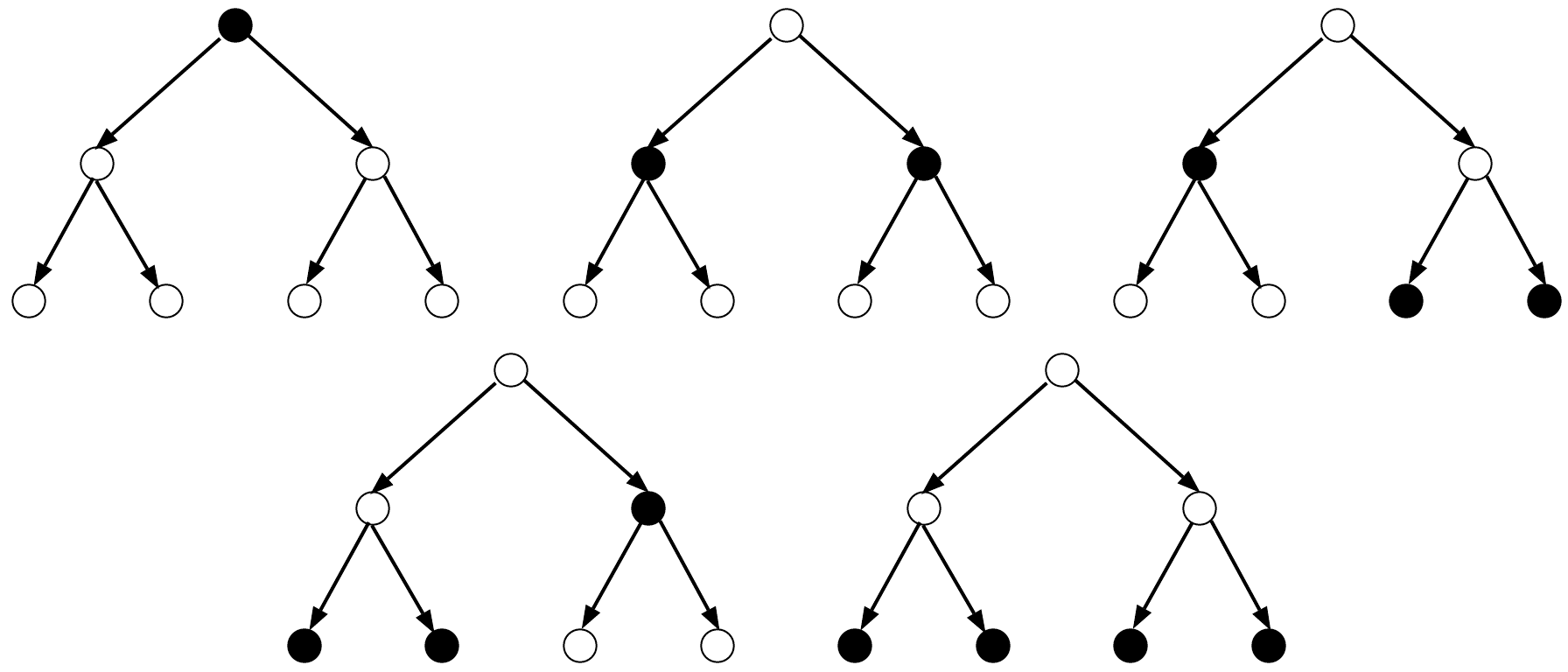}
\caption{Set of admissible activations for a binary tree of height $2$.}
\label{fig:activation}
\end{center}
\end{figure}

We now consider that the association of beams to flows $v_k$, $k \in \cK$ is fixed, in other words the flows are static. We determine which sets of beams can be activated simultaneously. To avoid interference, two beams $v$ and $v'$ may be activated at the same time if and only if they do not interfere. By definition of the codebook this is true if and only if $v$ is not a descendant or an ancestor of $v'$. We say that a subset of $V$ is an admissible activation strategy if all beams in this subset may be activated simultaneously without interfering with each other. We identify subsets of $V$ with vectors of $\{0,1\}^{|V|}$. Denote by $\bbz \in \{0,1\}^{|V|}$ a subset of beams where $z_v = 1$ if $v$ is activated and $z_v = 0$ otherwise. The set of admissible activation strategies is therefore:
$$
{\cal Z} = \left\{\bbz \in \{0,1\}^{|V|} :  z_v z_{v'} = 0 , \forall v' \in \cA(v), v \in V \right\}. 
$$
Figure~\ref{fig:activation} depicts the possible activation strategies when $G$ is a binary tree of height $2$. We denote the activated (resp. non-activated) beams by colored (resp. empty) vertices. It is noted that we solely depict the maximal activation strategies (i.e. for which activating a new beam would render the strategy non admissible).
Furthermore, we assume that one can perform time-sharing between the possible activation strategies, since in practice time is slotted and for each time slot one chooses an admissible set of beams to activate. Hence the vector of admissible proportions of time each beam is activated $\bbgamma$ must verify:
$$
\bbgamma = (\gamma_v)_{v \in V} \in \conv( \cal Z ).
$$
where $\conv$ denotes the convex hull, and $\gamma_v \in [0,1]$ is the proportion of time beam $v$ is activated. 

\subsection{Flow activation strategies}

When beam $v \in V$ is activated, at most one flow may be served by this beam at a time, and time sharing between flows associated to a given beam applies as well. Denote by $\delta_k \in [0,1]$ the proportion of time flow $k$ is served by beam $v_k$ divided by the proportion of time $v_k$ is activated. The admissible set for $\bbdelta=(\delta_v)_{v \in V}$ is hence:
$$
	\Delta = \left\{ \bbdelta \in [0,1]^{\cK}:  \sum_{k \in \cK(v)} \delta_k = 1 , v \in V \right\}. 
$$

\subsection{Fair rate allocations}

	We now determine fair rate allocation strategies, which involves finding the optimal values of $\bbgamma$ and $\bbdelta$ to maximize a utility function of the flows data rates. From the above, given $\bbgamma$ and $\bbdelta$, the effective data rate of flow $k \in \cK$ is given by $r_k \gamma_{v_k} \delta_k$.  We are interested in the family $\alpha$-fair allocations (see \cite{mo_fairendend_2000}), which leads to the following optimization problem: 
\als{
	\underset{{\bbdelta,\bbgamma}}{\text{Maximize }}  & \sum_{v \in V} \sum_{k \in \cK(v)} f_{\alpha}(r_k \gamma_{v} \delta_{k}) \\
	\text{   subject to   } & \gamma \in \conv(\cZ) \text{ and } \delta \in \Delta. 
}
where 
$$
f_{\alpha}(r) = \begin{cases} {r^{1 - \alpha} \over 1 - \alpha} & \text{ if } \alpha \ne 1 \\ \ln(r) & \text{ otherwise.} \end{cases}
$$
It is noted that this formulation includes Proportional Fairness (PF) ($\alpha = 1$), Maximum Throughput (MT) ($\alpha = 0$) and Max-Min Fairness (MMF) ($\alpha \to \infty$) as particular cases. Since this optimization problem is convex, one may find the optimal solution using an iterative method such as gradient descent or Newton's method, but we will show that, for any value of $\alpha$, an optimal solution may be computed by an efficient, problem specific, algorithm.

\subsection{Explicit expression for the convex hull}	
	Before deriving the optimal rate allocation, we state Proposition~\ref{prop:conv}, an auxiliary result which allows to express the convex hull $\conv({\cal Z})$ in closed form. The proof, which is relatively straightforward is omitted. The interpretation of Proposition~\ref{prop:conv} is that $\gamma_v$ can be written as the product $\kappa_v \prod_{v' \in \cA(v)}(1-\kappa_{v'})$, where $\kappa_v$ is the proportion of time $v$ is activated divided by the proportion of time none of its ancestors are activated. We recall that $v$ may be activated only if all of its ancestors are not activated.
	
	\begin{prop}\label{prop:conv}
		We have that:
		\als{
			\conv({\cal Z}) &= \left\{  \bbgamma: \gamma_v = \kappa_v \prod_{v' \in \cA(v)}(1-\kappa_{v'}),  \bbkappa \in  [0,1]^{|V|} \right\}\\
					            &= \left\{  \bbgamma \in [0,1]^{|V|}: \sum_{v' \in \bar{\cA}(v)} \gamma_{v'} \le 1, v \in V \right\}.
		}
	\end{prop}
	
\subsection{Efficient algorithms for fair rate allocation}

	We now derive the optimal rate allocation in closed form, and provide an efficient algorithm to compute it in practice.
\begin{theo}\label{th:alpha_fair_allocation}
Define the vectors $\bbphi,\bbtheta$ and $\bbkappa^\star$ as follows. Vector $\bbphi$ is given by:
$$
	\phi_v = \begin{cases} \left( \sum_{k \in \cK(v)} r_k^{ {1 \over \alpha} - 1} \right)^{\alpha} & \text{ if } \alpha \ne 1 \\
																 |\cK(v)| & \text{ otherwise.} \end{cases}
$$
Next define $\bbtheta$ and $\bbkappa$ through the following recursion. Let 
\begin{align*}
	\theta_v &= \begin{cases}  {\phi_v \over 1 - \alpha} & \text{if } v \text{ is a leaf} \\
	(\kappa^\star_v)^{1-\alpha} { \phi_v \over  1-\alpha} +  (1-\kappa^\star_v)^{1-\alpha} \sum_{v':(v,v') \in E} \theta_{v'} & \text{otherwise}.
	\end{cases} \\
	\kappa^\star_v &=  \left[ 1 + \left( {1-\alpha \over \phi_v}  \sum_{v':(v,v') \in E} \theta_{v'} \right)^{\frac{1}{\alpha}} \right]^{-1}.
\end{align*}
	Then $(\bbdelta^\star,\bbgamma^\star)$ the unique $\alpha$-fair allocation is given by:
\begin{align*}
	\delta_k^\star &= \begin{cases} {r_k^{{1 \over \alpha} - 1} \over \sum_{k' \in \cK(v_k)} r_{k'}^{ {1 \over \alpha} - 1}},& \text{ if } \alpha \ne 1 \\
																 {1 \over |\cK(v_k)|}, & \text{ otherwise.} \end{cases} \\
\gamma_v^\star &= \kappa^\star_v \prod_{v'\in \cA(v)}(1 - \kappa^\star_{v'}).  
\end{align*}
\end{theo}
The proof of Theorem~\ref{th:alpha_fair_allocation} is presented in Appendix. From Theorem~\ref{th:alpha_fair_allocation} we deduce Algorithm~\ref{algo:alpha_fair}, an efficient dynamic programming algorithm to calculate the $\alpha$-fair allocation given the flow rates $r_k$, $k \in \cK$ and the allocations $v_k$, $k \in \cK$. This algorithm runs in linear time ${\cO}(|V| + |\cK|)$ and is hence efficient, since its input has size ${\cO}(|V| + |\cK|)$.

\begin{algorithm}
\KwData{Tree $G$ sorted by decreasing height, flow rates $r_k$, $k \in \cK$, flow association $v_k$, $k \in \cK$, parameter $\alpha$}
 \tcc{Compute the aggregated data rates}
 \For{$v \in V$}{$\phi_v \gets 0$;}
 \For{$k \in \cK$}{$\phi_{v_k} \gets ( \phi_{v_k}^{1 \over \alpha} + r_k^{{1 \over \alpha} - 1})^{\alpha}$;}  
 \For{$k \in \cK$}{$\delta^\star_k \gets { r_k^{{1 \over \alpha} - 1} \over \phi_v^{1 \over \alpha}}$;}  
 \tcc{Dynamic Programming: ascending phase}
 \For{$v=|V|,|V|-1,\dots,1$}{
    $\tau_{v} \gets 0$;\\ 
		\For{$v': (v,v') \in E$}{
				$\tau_{v} \gets \tau_{v} + \theta_{v'}$;
				}
    $\kappa^\star_{v} \gets \left[ 1 + \left( {1-\alpha \over \phi_{v}} \tau_{v}  \right)^{\frac{1}{\alpha}} \right]^{-1}$;
    $\theta_v \gets (\kappa^\star_v)^{1-\alpha} { \phi_v\over  1-\alpha} +  (1-\kappa^\star_v)^{1-\alpha} \; \tau_{v}$;
    }
 \tcc{Dynamic Programming: descending phase}
 $\gamma^\star_{1} \gets \kappa^\star_1$;$h_{1} \gets 1- \kappa^\star_1$;\\
 \For{$v=2,\dots,|V|$}{
 	    $v' \gets $ father of $v$;\\
 	    $\gamma^\star_v \gets \kappa^\star_v h_{v'}$;\\
 	  	$h_{v} \gets h_{v'}(1 - \kappa^\star_v)$;\\
 	    }
 \KwResult{$\alpha$-fair allocation $\bbdelta^\star,\bbgamma^\star$}
 \caption{$\alpha$-fair allocation}
 \label{algo:alpha_fair}
\end{algorithm}

\subsection{Examples}

	We now illustrate the outcome of the $\alpha$-fair allocation for the important sub-cases of $\alpha \in \{0,1\}$. It is noted that the Max-Min Fair allocation (i.e. $\alpha \to \infty$) does not seem to yield a simple formula for arbitrary trees.
	
	{\bf Proportional Fairness ($\alpha = 1$)} The PF allocation is given by:
	\als{
		\delta^\star_k &= {1 \over n_{v_k}} \;,\; k \in \cK \\
		\kappa^\star_v &= {n_v  \over \sum_{v' \in \bar{\cD}(v)} n_{v'}} \;,\; v \in V
	}
	So that:
	$$
		\gamma^\star_v = {n_v  \over \sum_{v' \in \bar{\cD}(v)} n_{v'}} \prod_{v'' \in \cA(v)} {\sum_{v' \in \cD(v'')} n_{v'} \over \sum_{v' \in \bar{\cD}(v'')} n_{v'}}.
	$$
	While the proportion of time each beam is activated does not depend on the flow rates $r_k$, $k \in \cK$, the dependence in the number of flows served by each beam $\bbn = (n_v)_{v \in V}$ is not completely obvious.
	
	{\bf Maximal Throughput ($\alpha = 0$)} For simplicity, assume that all flows allocated to the same beam have the same data rate so that $r_k = r_{k'}$ if $v_{k} = v_{k'}$. Then we obtain:
	\als{
		\delta^\star_k &=  {1 \over n_{v_k}} \;,\;  k \in \cK \\
		\gamma^\star_v &= \prod_{v' \in \cD(v)} \indic\{  n_{v'} = 0 \} \;,\; v \in V. 
	}
	Hence a beam is activated all the time if all of its descendants are empty, and never activated otherwise. In short, Max Throughput gives absolute priority to beams with higher depth.
	
\begin{figure}[!h]
\begin{center}
  \subfloat[Example 1]{\includegraphics[width=0.8\columnwidth]{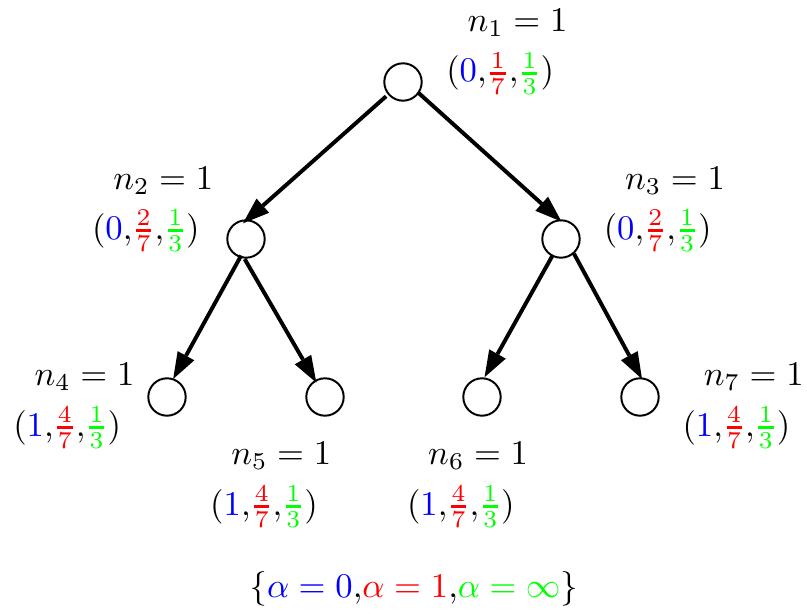}\label{fig:alpha_fairness1}} \\
  \subfloat[Example 2]{\includegraphics[width=0.8\columnwidth]{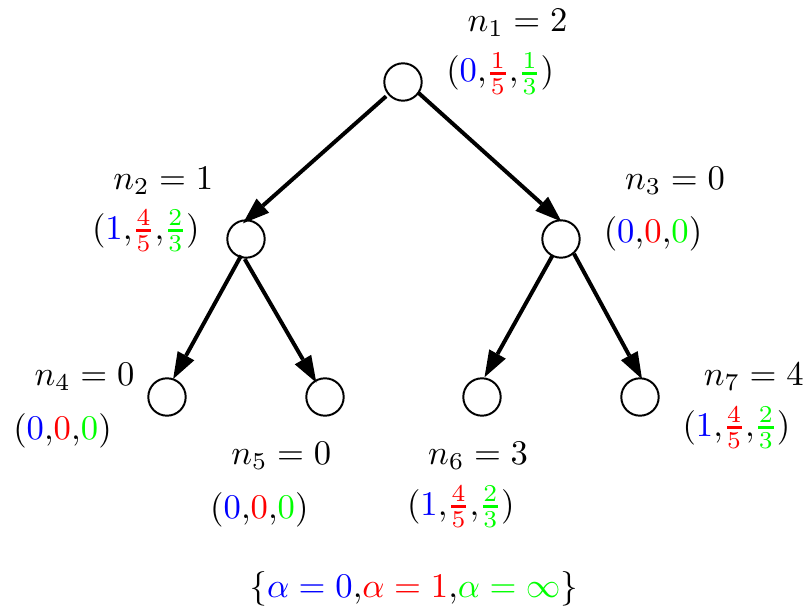}\label{fig:alpha_fairness2}}
\caption{Fair Rate Allocation}
\label{fig:alpha_fairness}
\end{center}
\end{figure}

	Figure~\ref{fig:alpha_fairness} illustrates the proportion of time each beam is activated by the $\alpha$-fair allocation on two simple examples. For simplicity we consider $G$ to be a binary tree of height $2$ and we assume that the data rate of each flow is unity, so that $r_k = 1$, $k \in \cK$. We can see that lower values of $\alpha$ tend to prioritize the leaves over the root, while higher values of $\alpha$ allocate more ressources to the root. Indeed, in the extreme case of $\alpha = 0$, the root is active if and only if all other nodes are empty. Also, as seen on Example $1$, the Proportional Fair allocation does not treat the root and the leaves equally if all of them have the same number of flows. This is due to the fact that activating the root prevents all other nodes from being activated.
			
\subsection{Practical implementation}

	At first sight, an algorithm which simply returns $\bbgamma^\star \in \conv({\cal Z})$ does not seem to be usable directly, since as said before, in practice time is slotted, and at each time slot one must select an admissible beam configuration from $\cZ$, so that the average proportion of time $v$ (over a large number of time slots) is activated equals $\gamma^\star_v$. In fact one may do so directly using a simple randomized algorithm. Given $\bbkappa^\star$ calculated by our algorithm, at each time slot, draw  ${\bf Y} = (Y_v)_{v \in V}$ independent Bernoulli random variables where $\EE({\bf Y}) = \bbkappa^\star$. Then define ${\bf Z} = (Z_v)_{v \in V} \in \{0,1\}^{|V|}$ as:
	$$
		Z_v = Y_v \prod_{v' \in \cA(v)} (1-Y_{v'}), v \in V
	$$
	and activate beam $v$ if and only if $Z_v = 1$. One may readily check that ${\bf Z} \in {\cal Z}$ since $Z_{v} Z_{v'} = 0$ whenever $v$ is an ancestor or a descendant of $v'$, and furthermore: 
	$$
	\EE(Z_v) = \kappa^\star_v \prod_{v' \in \cA(v)} (1-\kappa^\star_{v'}) = \gamma^\star_v , v \in V.
	$$
	Hence $\EE({\bf Z}) = \bbgamma^\star$, and repeating the above procedure for each time slot provides a practical implementation of the $\alpha$-fair allocation, and, as a bonus, involves calculating $\bbgamma^\star$ only once.

\section{Flow Level Performance for Elastic Traffic}\label{sec:elastic}

	We now study the flow-level performance of hierarchical beamforming for elastic traffic. Flows arrive at random time instants, want to download a file of exponentially distributed size and depart upon download completion. For each configuration of flows $\bbn \in \NN^{|V|}$, a given rate allocation scheme chooses a rate allocation $\bbdelta^\star(\bbn),\bbgamma^\star(\bbn)$, from which the throughput of each flow present in the system can be deduced. As done usually when studying the flow-level performance of wireless networks, we assume that the rate allocation mechanism adapts instantaneously upon arrival and departure of flows (separation of time scales), so that the throughput of any given flow at any given time is equal to the throughput given by the rate allocation scheme. Our aim is to calculate the throughput experienced by a typical flow. 
	
\subsection{Model}

	We assume that flows arrive in each region $\cG_v$ according to a Poisson process with rate $\lambda_v$. Each flow represents an amount of data whose size is exponentially distributed with expectation $1$. To simplify the analysis, we assume that any flow associated with beam $v$ has data rate $r_v$ (this can be relaxed in cases of interest, as shown below). We denote by $\bbN(t) = (N_v(t))_{v \in V}$ the state of the system at time $t$, which is a random variable, where $N_v(t)$ denotes the number of active flows present in $\cG_v$ at time $t$. Define the load for each beam:
$$
\rho_v = {\lambda_v \over r_v}
$$
It is noted that $\bbN(t)$ is a continuous-time Markov process. We say that the system is stable if and only if $\bbN(t)$ is positive recurrent. When the system is stable, Little's law guarantees that the expected amount of time a typical flow is served by beam $v$ equals ${\EE(N_v(t))\over \lambda_v} $. We then define the flow throughput (\cite{bonald_wirelessdownlinkdata_2003a}):
$$
	\psi_v = {\lambda_v \over \EE(N_v(t))},  
$$
which represents the expected throughput experienced by a typical flow served by beam $v$.
\subsection{Transition rates}
The transition rates of $\bbN(t)$ depend on the rate allocation mechanism. In state $\bbN(t)$, the proportion of time beam $v$ is activated equals $\gamma^\star_v(\bbN(t))$, so that the throughput of any flow served by beam $v$ equals 
$$r_v  \gamma^\star_v(\bbN(t)) \delta_{v}(\bbN(t)) =  {r_v \gamma^\star_v(\bbN(t)) \over N_v(t)}.$$
Denote by:
$$
\mu(\bbn \to \bbn') = {d \over d t} \PP( \bbN(t+dt) = \bbn' | \bbN(t) = \bbn)
$$
the transition rate of the Markov process $\bbN(t)$ between states $\bbn$ and $\bbn'$. Denote by $\bbe^v = (0,\dots,0,1,0,\dots,0)$ the $v$ th-canonical base vector of $\RR^{|V|}$. The only possible transitions from $\bbn$ are $\bbn \to \bbn + \bbe^v$ and $\bbn \to \bbn - \bbe^v$, $v \in V$, which correspond to an arrival and a departure of a flow served by beam $v$ respectively. The transition rates are:
\begin{align*}
	\mu(\bbn \to \bbn + \bbe^v) &= \lambda_v, \\
	\mu(\bbn \to \bbn - \bbe^v) &= {r_v \gamma^\star_v(\bbn)} \indic\{ n_v \ge 1\}.
\end{align*}

\subsection{Stability region}
The stability of the system depends on the rate allocation mechanism which maps the system state $N(t)$ into the corresponding rate allocation $\bbdelta(\bbN(t)), \bbgamma(\bbN(t))$. We may derive the stability region from our earlier analysis of fair scheduling.
\begin{prop}\label{prop:stability}
	There exists a rate allocation scheme ensuring stability if and only if $\sum_{v' \in \bar{\cA}(v)} \rho_{v'} <  1$ for all $v \in V$. 
\end{prop}
\bp There exists a rate allocation scheme ensuring stability if and only if $\bbrho$ lies in the interior of  $\conv(\cZ)$. Furthermore, from Proposition~\ref{prop:conv} $$\conv(\cZ) = \left\{  \bbgamma \in [0,1]^{|V|}: \sum_{v' \in \bar{\cA}(v)} \gamma_{v'} \le 1, v \in V \right\}$$ which concludes the proof. \ep

\section{Flow level performance under proportional fairness}\label{sec:elastic_pf}

\subsection{PF allocation}
We now compute the stationary distribution of $\bbN(t)$ for Proportional Fair rate allocation which is the $\alpha$-fair rate allocation scheme described in the previous section with $\alpha = 1$. We recall that the PF allocation is given by:
\begin{align*}
	\gamma^\star_v(\bbn) &= {n_v  \over \sum_{v' \in \bar{\cD}(v)} n_{v'}} \prod_{v'' \in \cA(c)} {\sum_{c' \in \cD(v'')} n_{v'} \over \sum_{v' \in \bar{\cD}(v'')} n_{v'}} \\
	\delta^\star_v(\bbn) &= {1 \over n_v}.
\end{align*}

{\bf Stationary Distribution and Flow Throughput} Both the stationary distribution and the flow throughput are given by Theorem~\ref{th:pf_flow}. The proof of Theorem~\ref{th:pf_flow} is based on the fact that $\bbN(t)$ is reversible, so that its stationary distribution is known up to a normalization constant, and then calculating this constant by recursion. It should be noted that, in general, the PF allocation does not lead to reversible systems. It is also noted that the result holds not only for exponential flow size distributions, but in fact holds for all flow sizes distributions i.e. the system is insensitive.
\begin{theo}\label{th:pf_flow}
Consider PF rate allocation. Then $\bbN(t)$ is a reversible Markov process with stationary distribution:
$$
\pi(\bbn) = {1 \over c(\bbrho)} \prod_{v \in V} \rho_v^{n_v} {\sum_{v' \in \bar{\cD}(v)} n_{v'} \choose n_v} 
$$
with:
$$
c(\bbrho) =\prod_{v \in V} {1-\sum_{v'\in \cA(v)} \rho_{v'} \over 1-\sum_{v' \in \bar{\cA}(v)}\rho_{v'}}.
$$
The expected number of customers in stationary state is:
$$
\mathbb{E}[N_v(t)] = \rho_{v} \sum_{v' \in \bar{\cD}(v)} {1-\deg(v') \over 1-\sum_{v''\in \bar{\cA}(v')} \rho_{v''}}
$$
and the flow throughput is:
$$
 \psi_v = r_v \left[\sum_{v' \in  \bar{\cD}(v)} {1-\deg(v') \over 1-\sum_{v''\in \bar{\cA}(v')}\rho_{v''}} \right]^{-1}
$$
Furthermore, the above holds for all flow time distributions, i.e. the system is insensitive.
\end{theo}
\subsection{Proof of Theorem~\ref{th:pf_flow}}
To prove that $\bbN(t)$ is both reversible and has stationary distribution $\pi$, it is sufficient to check that the detailed balance condition holds:
$$
	\pi(\bbn) \mu(\bbn \to \bbn + \bbe^{v}) = \pi(\bbn + \bbe^{v}) \mu(\bbn + \bbe^{v} \to \bbn)
$$	
For all $\bbn \in \NN^{|V|}$ and $v \in V$. Detailed balance holds by inspection since:
$$
	{\mu(\bbn \to \bbn + \bbe^{v}) \over  \mu(\bbn + \bbe^{v} \to \bbn)}  = {\rho_v \over \gamma^\star_v(\bbn + \bbe^{v})} = {\pi(\bbn + \bbe^{v}) \over \pi(\bbn)}.
$$
Hence $\bbN(t)$ has stationary distribution $\pi$ and the value of $c(\bbrho)$ can be found by normalization since $\sum_{\bbn \in \NN^{|V|}} \pi(\bbn) =1$:
$$
	c(\bbrho) = \sum_{\bbn \in \NN^{|V|}} \prod_{v \in V}  \rho_v^{n_v} {\sum_{v' \in \bar{\cD}(v)} n_{v'} \choose n_v}.
$$
We proceed by recursion, and for $v \in V$, we define $c^v(\bbrho)$ the value of $c(\bbrho)$ when $\rho_{v'} = 0$ for all $v' \not\in \bar{\cD}_{v}$, i.e. $c^v(\bbrho)$ is the value of the above sum only considering $\bbn \in \NN$ with $n_{v'} = 0$ unless $v' \not\in \bar{\cD}_{v}$. Another interpretation is that $c^v(\bbrho)$ is the value of $c(\bbrho)$ when considering the subgraph made of $v$ and its descendants. Finally it is also noted that $c(\bbrho) = c^{1}(\bbrho)$ since $1$ is the root. We use the following fact for binomial coefficients: 
\begin{fact} \label{fact:binom} 
	For $z,\beta \in \CC$, we have $ {1 \over (1 - z)^{\beta + 1}} = \sum_{k \ge 0} {k + \beta \choose k} z^k$. 
	\end{fact}
Summing over $n_1 \in \NN$ the above expression we obtain the following relation:
$$
	c(\bbrho) = c^1(\bbrho) = {1 \over 1 - \rho_1} \prod_{v: (1,v) \in E} c^v\left({\bbrho \over 1- \rho_1}\right).  
$$
More generally we have the following recursive relation:
$$
	c^v(\bbrho) = {1 \over 1 - \rho_v} \prod_{v': (v,v') \in E} c^{v'}\left({\bbrho \over 1- \rho_v}\right).
$$
Iterating the relation above starting at $v=1$ we obtain the announced result:
$$
	c(\bbrho) = \prod_{v \in V} {1-\sum_{v'\in \cA(v)}\rho_{v'} \over 1-\sum_{v' \in \bar{\cA}(v)}\rho_{v'}}.
$$
The expected number of customers is calculated using the following trick. Recall the definition of $c(\rho)$:
$$
	c(\bbrho) = \sum_{\bbn \in \NN^{|V|}} \prod_{v' \in V} \rho_{v'}^{n_{v'}} {\sum_{v'' \in \bar{\cD}(v')} n_{v''} \choose n_{v'}}
$$
Taking logarithms and differentiating with respect to $\rho_v$:
\als{
	\rho_v {\partial \ln c(\rho) \over \partial \rho_v} &= {1 \over c(\rho)} \sum_{n \in \NN^{|V|}} n_v \prod_{v' \in V} \rho_{v'}^{n_{v'}} {\sum_{v'' \in \bar{\cD}(v')} n_{v''} \choose n_{v'}} \\
	&= \sum_{n \in \NN^{|V|}} n_v \pi(\bbn) = \EE(N_v(t)).
}
Plugging the previous expression of $c(\rho)$ gives:
\als{
	\EE(N_v(t)) &= \rho_v {\partial \ln c(\rho) \over \partial \rho_v} \\
	&= \rho_v {\partial \over \partial \rho_v} \sum_{v' \in V}  \ln\left( {1-\sum_{v''\in \cA(v')}\rho_{v''} \over 1-\sum_{v'' \in \bar{\cA}(v')}\rho_{v''}}\right) \\
	&= \rho_{v} \sum_{v' \in  \bar{\cD}(v)} {1-\deg(v') \over 1-\sum_{v''\in \bar{\cA}(v')} \rho_{v''}}.
}
as announced. 

\section{Flow level performance under maximal throughput}\label{sec:elastic_mt}	
	
We now study the flow-level performance for the Maximal Throughput rate allocation which is the $\alpha$-fair rate allocation scheme with $\alpha = 0$. As we can see, while in this case $\bbN(t)$ is not reversible (nor insensitive), it has a hiearchical structure which can be exploited in order to provide tractable expressions for performance. 

{\bf Transition Rates} We first calculate the transition rates. From the hierarchical nature of the codebook, we have that $r_{v'} > r_v$ if $v' \in D(v)$ (see Definition~\ref{def:codebook}). As stated in the previous section, in state $\bbn \in \NN^{|V|}$, the proportion of time beam $v$ is activated equals:
	\als{
		\gamma^\star_v(\bbn) &= \prod_{v' \in \cD(v)} \indic\{  n_{v'} = 0 \}. 
	}
	namely, beam $v$ is activated all the time if all of its descendants are empty, and never activated otherwise. 	

\subsection{Performance for line graphs}\label{subsec:LG}

We first consider the case of line graphs, where we have $E = \{(1,2),(2,3),\dots,(|V|-1,|V|)\}$. From definition~\ref{def:codebook}, beam $v$ is activated at time $t$ if and only if all of its descendants are empty, namely $n_{v'}(t) = 0$  $v'=v+1,\dots,|V|$. From this observation, we can reduce the system to a M/M/1 queue with preemptive-resume priority as follows. Indeed, consider a single server with $|V|$ classes of users, where class $v$ represents users served by beam $v$. In this system, at time $t$, users of class $v$ are served at rate $r_v$ if $n_{v'}(t) = 0$, $v'=v+1,\dots,|V|$ and at rate $0$ otherwise. This new system is equivalent to the original system, which yields Theorem~\ref{th:mt_line}. The proof is provided in appendix~\ref{sec:proof_th_mt_line}.

\begin{theo}\label{th:mt_line}
Consider MT allocation and $G$ a line graph. Then $\bbN(t)$ is positive recurrent (stable) if and only if $	\sum_{v \in V} \rho_{v} <  1$. Furthermore, if $\sum_{v \in V} \rho_{v} <  1$ we have for all $v \in V$:
$$	
	\EE(N_v(t)) = \frac{\rho_v\left(1+\sum_{v' \ge v}\rho_{v'}\left(\frac{r_v}{r_{v'}}-1\right)\right)}{ \left(1-\sum_{v' \ge v }\rho_{v'}\right)\left(1-\sum_{v' > v} \rho_{v'}\right)}   
$$
and the flow throughput is :\\
$$
\psi_v = \frac{ r_v (1-\sum_{v' \ge v}\rho_{v'})(1-\sum_{v > v'}\rho_{v'})}{ 1+\sum_{v'\ge v}\rho_{v'}(\frac{r_v}{r_{v'}}-1)}
$$
\end{theo}

\subsection{Performance for generic graphs}

We now consider the case where $G$ is a generic tree. In that case as well we compute the flow throughput and the analysis is much more involved than for the previous case. Under the MT allocation, a beam is activated if and only if all of its descendants are empty. However the state of a beam does not depend on the state of its ancestors. Therefore, the evolution of $\bbN_v(t)$ depends on $\bbN_{v'}(t)$ if and only if $v' \in \bar{\cD}(v)$. In order to study the distribution of the state of $v$, i.e. $N_v(t)$, it is natural to study the process $(N_v(t))_{v' \in \bar{\cD}(v)}$ which describes its state and that of its descendants. We define the busy period of this process which plays a crucial role in our analysis. We define $\ell_v = \sum_{v' \in \cD(v)} \lambda_{v'}$ the total arrival rate in the descendants of $v$.

\begin{defin}\label{def:cicle_times}
For $v \in V$ define the Markov processes 
\begin{align*}
\bbN^v(t) &= (N_{v'}(t))_{v' \in \bar{\cD}(v)} \in \NN^{|\bar{\cD}(v)|}, \\
\bbM^v(t) &= (N_{v'}(t))_{v' \in \cD(v)} \in \NN^{|\cD(v)|},
\end{align*}
where $\bbN^v(t)$ describes the state of $v$ and its descendants, and $\bbM^v(t)$ describes the state of the descendants of $v$. Define the random variable $B_v$ which is the busy period of process $\bbM^v(t)$. We recall that the busy period is $B_v = T_2 - T_1$ in a system where $\bbM^v(0) = {\bf 0}$, and $T_1 > 0$ is the first random instant after $0$ such that $\bbM^v(T_1) \ne {\bf 0}$, and $T_2 > T_1$ is the first random instant after $T_1$ such that $\bbM^v(T_2) = {\bf 0}$. 
\end{defin}
In Theorem~\ref{th:mean}, we show that the expected number of customers in stationary state can be computed as a function of the two first moments of the busy period.
\begin{theo}\label{th:mean}	
Under MT rate allocation, the number expected number of flows in beam $v$ satisfies, for all $v \in V$:
$$
\EE(N_v(t)) = {  {\lambda_ v\ell_v \over 2} \EE(B_v^2) + \rho_v \left(1 + \EE(B_v) \ell_v \right)^2  \over (1 - \rho_v(1 + \EE(B_v) \ell_v ))(1 + \EE(B_v) \ell_v) }
$$
Furthermore, $\EE(B_v)$ can be computed recursively using the following relations:
$$
	\EE(B_v) = {1 \over \ell_v}\left( {1 \over  \prod_{v':(v,v')\in E} \PP( \bbN^{v'}(t) = 0)} - 1\right).
$$
and:
\als{
	\PP(\bbN^v(t) = 0) +\rho_v = \prod_{v':(v,v') \in E} \PP(\bbN^v(t) = 0), \forall v \in V.
}
\end{theo}
	
\subsection{Proof of Theorem~\ref{th:mean}}

{\bf Expected number of customers.} We say that a customer is "waiting" if it has never received service, and it is "in service" if it has received service but has not left the system yet.  We compute the expected time spent in service, denoted by $\EE(M_v)$. When a flow first receives service by beam $v$, by definition, no flows from classes $v'\in \cD(v)$ must be present in the system. Once she has started to receive service, there are two possible outcomes: either no flows from classes $v'\in \cD(v)$ arrive before her service completion (this occurs with probability ${r_v \over r_v + \ell_v}$), otherwise she must wait for the duration of a busy period of $\bbM^v(t)$ and we are back in the initial situation, since the exponential distribution is memoryless (this occurs with probability ${ \ell_v \over r_v + \ell_v}$). The expected time before either of these events happens is ${1 \over r_v + \ell_v}$. Therefore the expected service time verifies:
$$
	\EE(M_v) = {r_v \over r_v + \ell_v} {1 \over r_v} +  {\ell_v \over r_v + \ell_v}( \EE(B_v) + \EE(M_v))
$$
so that:
$$
	\EE(M_v) = {1 \over r_v} (1 + \EE(B_v) \ell_v)
$$
We now calculate the expected time spent waiting denoted by $\EE(W_v)$. Assume that a flow arrives at time $0$. From the PASTA property, $N_v(0)$ is distributed as the stationary distribution. Denote by $T_v^{-}$ the first instant at which $\bbM^v(T_v^-) = 0$. Denote by $T_v^{-} + T_v^{+}$ the first instant at which all customers present at time $0$ have been served. Therefore the waiting time may be decomposed as: 
$$
	\EE(W_v) = \EE(T_v^{-}) + \EE(T_v^{+}).
$$
First one notices that $T_v^{-}$ is expressed as a function of the residual busy period of the process $\bbM^v(t)$ and is given by:
$$
	\EE(T_v^{-}) = {1 \over 2} {\EE(B_v^2) \over \EE(B_v) + \EE(A_v)} = {1 \over 2} {\EE(B_v^2) \over \EE(B_v) + \ell_v^{-1}}.
$$
where $A_v$ is the time that the process $\bbM^v(t)$ spends in state $\bf 0$. Indeed $A_v$ is exponentially distributed with mean ${1 \over \ell_v}$.

From Little's law, $\EE(N_v(0)) = \lambda_v(\EE(W_v) + \EE(M_v))$. No flow of beam $v$ may receive service before $T_v^-$, therefore $T_v^+$ is the amount of time necessary to serve $N_v(0)$ flows and is given by:
$$
	\EE(T_v^+) = \EE(N_v(0)) \EE(M_v) = \lambda_v \EE(M_v) (\EE(W_v) + \EE(M_v)) .
$$
using the previous relation. Substituting we obtain:
\als{
	\EE(W_v) = \EE(T_v^{-}) + \EE(T_v^{+}) 
					 = \EE(T_v^{-}) + \lambda_v \EE(M_v) (\EE(W_v) + \EE(M_v)),
}
which yields:
$$
	\EE(W_v) = {\EE(T_v^{-}) + \lambda_v \EE(M_v)^2 \over 1 - \lambda_v \EE(M_v)} 
$$	
Therefore the total expected time spent by a typical flow is given by:
\als{
	\EE(W_v) + \EE(M_v) &=  {\EE(T_v^{-}) + \EE(M_v) \over 1 - \lambda_v \EE(M_v)}
}	
Substituting the value of $\EE(M_v)$ we get
\als{
 	\EE(W_v) + \EE(M_v) &= {  {1 \over 2} {\EE(B_v^2) \over \EE(B_v) + \ell_v^{-1}} + {1 \over r_v} \left(1 + \EE(B_v) \ell_v \right)  \over 1 - \rho_v(1 + \EE(B_v) \ell_v )  } \\
 	&= {  {\ell_v \over 2} \EE(B_v^2) + {1 \over r_v} \left(1 + \EE(B_v) \ell_v \right)^2  \over (1 - \rho_v(1 + \EE(B_v) \ell_v ))(1 + \EE(B_v) \ell_v) }.
}
Using Little's law $\EE(N_v(t)) = \lambda_v(\EE(W_v) + \EE(M_v))$ yields the announced result.

{\bf Expected duration of the busy periods}. Consider the Markov process $\bbM^v(t)$. The expected duration this process spends in state $\bf 0$ is ${1 \over \ell_v}$, and the duration between two visits to state $\bf 0$ equals $\EE(B_v)$. Therefore, the probability that $\bbM^v(t)$ is in state $\bf 0$ satisfies:
$$
	\PP(\bbM^v(t) = 0) = {  {1 \over \ell_v} \over {1 \over \ell_v} + \EE(B_v) },
$$
so that the expected busy period is:
$$
	\EE(B_v) = {1 \over \ell_v}\left( {1 \over \PP(\bbM^v(t) = 0)} - 1\right).
$$
Furthermore, using the fact that, if $v'$ and $v''$ are distinct children of $v$, processes $\bbN^{v'}(t)$ and $\bbN^{v''}(t)$ are independent we get:
\als{
	\PP(\bbM^v(t) = 0) &= \PP( \bbN^{v'}(t) = 0, \forall v':(v,v') \in E) \\
	&= \prod_{v':(v,v')\in E} \PP( \bbN^{v'}(t) = 0).
}
Substituting yields the second claim:
$$
	\EE(B_v) = {1 \over \ell_v}\left( {1 \over  \prod_{v':(v,v')\in E} \PP( \bbN^{v'}(t) = 0)} - 1\right).
$$

{\bf Void probabilities} To complete the proof, we must compute the void probabilities $\PP(\bbN^{v}(t) = 0)$ for $v \in V$. We have:
\als{
	\PP( \bbM(t)^v = 0)  = \PP( \bbM(t)^v = 0 , N_v(t) \ne 0) + \PP( \bbM(t)^v = 0 , N_v(t) = 0) 
}
As explained before, $\PP( \bbM(t)^v = 0) = \prod_{v':(v,v')\in E} \PP( \bbN^{v'}(t) = 0) $ from indpendence, and by definition $\PP( \bbM(t)^v = 0 , N_v(t) = 0) = \PP( \bbN(t)^v = 0)$. Finally, beam $v$ serves its users if and only if $\PP(\bbM(t)^v = 0 , N_v(t) \ne 0)$, so that, by conservation of work we must have $\lambda_v = r_v \PP(\bbM(t)^v = 0 , N_v(t) \ne 0)$. Substituting yields the announced relation:
\als{
	\prod_{v':(v,v')\in E} \PP( \bbN^{v'}(t) = 0) =  \rho_v + \PP( \bbN(t)^v = 0) 
}
and from this relation, the value of $\PP( \bbN^{v}(t) = 0)$ may be computed by backwards induction for all $v \in V$ which concludes our proof.

\subsection{Busy periods: exponential approximation}\label{subsec:mt_height2_approx}

	As shown by Theorem~\ref{th:mean}, the only missing piece in order to compute $\EE(N_v(t))$ is the value of the second moment of the cycle times $\EE(B_v^2)$.  In general, computing the second moment of the busy period does not seem completely straightforward. We propose to approximate the busy period by an exponential, so that ${\EE(B_v^2) \over 2} \approx  \EE(B_v)^2$. The flow throughput is given by corollary~\ref{th:mt_h2}. The rationale for this approximation is as follows: if $\bbrho \ll {\bf 1}$, then the busy period is simply the service time of the first customer which enters the system, and this time indeed has exponential distribution. As shown in our numerical experiments, this approximation yields tractable formulas which are rather accurate.
	
\begin{corollary}\label{th:mt_h2}
Assume that the busy period can be approximated by an exponential. Then, under MT rate allocation, the number expected number of flows in beam $v$ satisfies, for all $v \in V$:
$$
\EE(N_v(t)) = {  \lambda_ v\ell_v \EE(B_v)^2 + \rho_v \left(1 + \EE(B_v) \ell_v \right)^2  \over (1 - \rho_v(1 + \EE(B_v) \ell_v ))(1 + \EE(B_v) \ell_v) }
$$
\end{corollary}

\section{Flow Level Performance for Streaming Traffic}\label{sec:streaming}
\subsection{Model} 

We now consider streaming traffic where flows require a fixed amount of resources throughout their stay in the system. Namely the available resources are spit into an integer number $\xi \ge 1$ of circuits, where a circuit represents the smallest unit of resource that can be allocated to a flow. Each flow served by beam $v \in V$ requires a number of circuits $s_v \in \{1,\dots,\xi\}$, and to ensure that that all flows are allocated enough resources admission control is used. When a new flow arrives, one checks whether or not one can guarantee that all active flows can be allocated enough circuits. If the answer is yes, the flow is admitted, otherwise she is blocked. Flows arrive in beam $v$ according to a Poisson process with rate $\lambda_v$ and remain there during an exponentially distributed time with mean ${1 \over r_v}$. Define the load $\rho_v={\lambda_v \over r_v}$. In fact the system is insensitive, so that considering exponential service times is sufficient. The state of the system at time $t$ is $\bbN(t) \in \cN \subset \NN^{|V|}$ where $\bbN_v(t)$ represents the number of flows served by beam $v$ and $\cN$ is the set of states in which all active flows can be allocated enough circuits. The only possible transitions of $N(t)$ correspond to the arrival and departure of a flow, and the transition rates are for all $\bbn \in \cN$:
\begin{align*}
	\mu(\bbn \to \bbn + \bbe^v) &= \lambda_v \indic\{ \bbn + \bbe^v \in \cN\}, \\
	\mu(\bbn \to \bbn - \bbe^v) &= r_v \indic\{ n_v \ge 1\}.
\end{align*}

\subsection{Admission control} 

	Let us now characterize the set of admissible states allowed by admission control i.e. states in which all flows can be allocated enough circuits by some policy. Consider the system in state $\bbn \in \NN^{|V|}$, and assume that there exists a rate allocation policy $\bbgamma^\star(\bbn) \in \conv({\cZ})$ which ensures that all flows are allocated enough circuits. Flows served by beam $v$ require $s_v$ circuits so that the proportion of time $v$ is activated should satisfy $\bbgamma^\star_v(\bbn) = n_v {s_v \over \xi}$ for all $v$. Furthermore, since any feasible rate allocation policy must satisfy $\bbgamma(\bbn) \in \conv(\cZ)$. Clearly, both of those conditions can be satisfied if and only if:
	$$
		 \sum_{v' \in \bar{\cA}(v)} n_{v'} {s_{v'} \over \xi} \le 1, v \in V.
	$$
The set of admissible states $\cN$ is therefore:
$$
\cN = \left\{ \bbn \in \NN^{|V|} : \sum_{v' \in \bar{\cA}(v)} n_{v'} s_{v'}  \le \xi \, , \,  v \in V \right\}.
$$
Furthermore, if the system is in state $\bbn \in \cN$, and a flow associated with beam $v$ arrives, it is admitted if and only if his entering the system leads to an admissible state, and blocked otherwise. Define the set of blocking states for beam $v$: 
$$
\cN(v) = \{ \bbn \in \NN^{|V|} : \bbn \in \cN, \bbn + e^v \not\in \cN \}.
$$
It is noted that a flow associated with beam $v$ will be blocked if and only if it arrives in a blocking state for beam $v$, that is  $\bbn \in \cN(v)$.
\subsection{Stationary distribution and blocking probability}  

	We may now derive the stationary distribution of the system in closed form, as well as the blocking probability for each beam, which constitutes the performance figure of the system. The stationary distribution is known in closed form up to a normalization constant, so that the main difficulty is to compute a sum over the state space $\cN$. Clearly, brute force summation is not feasible as $|\cN|$ grows exponentially with $|V|$. We show that there exists a low complexity algorithm (stated as Algorithm~\ref{algo:streaming}) to compute the stationary distribution as well as blocking probabilities. The result is stated as Theorem~\ref{th:streaming} and proven in subsection~\ref{subsec:proof_theorem_streaming}.
	
	It is noted that Algorithm~\ref{algo:streaming} computes the blocking probabilities in time $\cO( \xi |V| h(G))$. In most cases of interest, for instance if $G$ is a regular tree, $h(G) = \cO(\ln |V|)$, so that the running time of Algorithm~\ref{algo:streaming} is $\cO( \xi |V| \ln |V| )$ which is linear in the number of circuits $\xi$ and almost linear in the number of beams $|V|$. Since its input has size $\cO(|V|)$, we deduce that the dependency of the running time of this algorithm on $|V|$ is optimal up to a logarithmic factor $\ln |V|$. This algorithm is reminiscent of the Kaufman-Roberts algorithm~\cite{kauffman_1981,roberts_1981} used to compute blocking probabilities in multi-rate Erlang systems, and leverages the hierchical structure of the codebook by computing the blocking probabilities recursively. 

\begin{theo}\label{th:streaming}
	(i) Under the above assumptions, $\bbN(t)$ is a reversible Markov process with stationary distribution $\pi$:
$$
  \pi(\bbn) = {1 \over c(\bbrho)} \prod_{v \in V} {\rho_v^{n_v} \over n_v!} \,\, , \, \, \bbn \in \cN,
$$
where $c(\rho)$ is the normalization constant defined as:
$$
	c(\bbrho) = \sum_{\bbn \in \cN} \prod_{v \in V} {\rho_v^{n_b} \over n_b!}.
$$
(ii) The blocking probability of class $v$ is:
$$
	p_v(\bbrho) = \PP( \bbN(t) \in \cN(v)) = {\sum_{n \in \cN(v)} \prod_{v' \in V} {\rho_{v'}^{n_{v'}} \over n_{v'}!} \over \sum_{n \in \cN} \prod_{v' \in V} {\rho_{v'}^{n_{v'}} \over n_{v'}!}}.
$$ 
(iii) Algorithm~\ref{algo:streaming} outputs of value of the blocking probability vector $\bbp(\bbrho)$ using time $\cO( \xi |V| h(G))$ and memory $\cO(|V| \xi)$.
\end{theo}

\begin{algorithm}
\KwData{Tree $G$ sorted by decreasing height, loads $\bbrho$, circuit requirements $\bbs$, number of circuits $\xi$}  
\textbf{for} {$v=1,\dots,|V|$ \textbf{\textit{and}} $s=1,\dots,\xi$ \textbf{do}  $c_v(s) \gets 0$;\\}
 \tcc{Phase 1: Compute the normalization constant}
\For {$v=|V|,\dots,1$ \textbf{and} $s=1,\dots,\xi$}{
 \For {$\ell=0,\dots,\floor{s /s_v}$}{
 $t \gets 1$;\\
 \textbf{for}{ $v': (v,v') \in E$ \textbf{do} $t \gets t c_{v'}(s - \ell s_v)$;\\}
 $c_v(s) \gets c_v(s) + {\rho_v^{\ell} \over \ell!} t$;\\
 }
 \textbf{for} {$v'=|V|,\dots,1$ \textbf{do} $q_{v,v'}(s) \gets c_{v'}(s)$;\\}}
\tcc{Phase 2: compute blocking probabilities}
 \For {$v=1,\dots,|V|$ \textbf{and} $s=1,\dots,\xi$}{
 $v' \gets v$;\\
 \textbf{if} {$s > s_v $ \textbf{do} $q_{v,v'}(s) \gets c_v(s - s_v)$;\\}
 \While {$v'\neq 1$}{
 $v' \gets $ father of $v'$;\\  
 $q_{v,v'}(s) \gets 0$;\\
 \For {$\ell=0,\dots,\floor{s /s_{v'}}$}{
 $t \gets 1$;\\
 \textbf{for}{ $v'': (v',v'') \in E$ \textbf{do} $t \gets t q_{v,v''}(s - \ell s_{v'})$;}
 $q_{v,v'}(s) \gets q_{v,v'}(s) + {\rho_{v'}^{\ell} \over \ell!} t$;\\
 }	
 	}
$p_v \gets 1 - { 	q_{v,1}(\xi) \over c_1(\xi)}$;\\
}
 \KwResult{Blocking probabilities $\bbp$}
 \caption{Computation of blocking probabilities}\label{algo:streaming}
\end{algorithm}

\subsection{Proof of Theorem~\ref{th:streaming}}\label{subsec:proof_theorem_streaming}

{\bf Stationary Distribution.} By, inspection, one may readily check that $\pi$ verifies the detailed balance conditions:
$$
	{\pi(\bbn + \bbe^v) \over \pi(\bbn)} = {\mu(\bbn \to \bbn + \bbe^v) \over \mu(\bbn + \bbe^v \to \bbn)} = {\lambda_v \over r_v} = \rho_v.  
$$
so that $\bbN(t)$ is indeed reversible with stationary distribution $\pi$. 

{\bf Normalization Constant} It now remains to compute the normalization constat $c(\bbrho)$. Furthermore, for $s=0,1,\dots,\xi$, and $v \in V$ define $\cN(s,v)$ the set of admissible sets when there are $s$ circuits, and where beams which do not descend from $v$ are empty:
\begin{align*}
\cM(s,v) = \Big\{ \bbn \in \NN^{|V|} &: \sum_{v'' \in \bar{\cA}(v')} n_{v''} s_{v''}  \le s, \forall v' \in V \\
 &\text{ and }  n_{v'} = 0, \forall v' \not\in \bar{\cD}(v)  \Big\}.
\end{align*}
and we define the corresponding partial sums:
$$
	c_v(s) = \sum_{\bbn \in \cM(s,v)} \prod_{v' \in \bar{\cD}(v)} {\rho_v^{n_{v'}} \over n_{v'}!}.
$$
It is noted that $\cM(\xi,1) = \cN$, so that $c = c_1(\xi)$. The idea of the proposed algorithm is to compute the value of $c_v(s)$ for $v \in V$ and $s=0,\dots,\xi$ by starting at the leaves and then using backward induction.
Consider the subtree rooted at $v$ and $v'$ such that $(v,v') \in E$. If there are $s$ circuits and beam $v$ is serving $n_v$ users, $s_v n_v$ circuits need to be allocated to beam $v$, so that the number of circuits available to all the beams in the subtree rooted at $v'$ is $s - s_v n_v$. Therefore, $\bbn \in \cM(s,v)$ if and only if it admits the following decompostion:
$$
		\bbn = n_v \bbe^v + \sum_{v':(v,v') \in E} \bbm^{v'} \text{ with } \bbm^{v'} \in \cM(v',s-s_v n_v),  
$$
This fact gives rise to the following recursion, by summing over $n_v$ in the definition of $c_v(s)$:
\als{
	c_v(s) = \displaystyle \sum_{n_v=0}^{\floor{{s  \over s_v}}} {\rho_v^{n_v} \over n_v!} \prod_{v':(v,v') \in E} c_{v'}(s - n_v s_v).
}
The above relation readily gives an algorithm to compute $c_v(s)$, for all $v \in V$ and $s \le \xi$ which consistutes the first part of our algorithm. 
  
{\bf Blocking Probabilities} As said before, the blocking probability $p_v$ is the probability for a flow to arrive at a time where the system is in a blocking state $\bbn \in \cN(v)$,  so that the blocking rate for class $v$ is:
$$
p_v = \PP( \bbN(t) \in \cN(v)) = {\sum_{\bbn \in \cN(v)} \prod_{v' \in V} {\rho_{v'}^{n_{v'}} \over n_{v'}!} \over \sum_{\bbn \in \cN} \prod_{v' \in V} {\rho_{v'}^{n_{v'}} \over n_{v'}!}}.
$$ 
We compute the blocking rate using a similar approach as that used above for the normalization constant.  We have that $\bbn \in \cN \setminus \cN(v)$ if and only if $\bbn + \bbe^v \in \cN$, which, by definition of $\cN$ translates to:
$$
	\sum_{v' \in \bar{\cA}(v'')} n_{v'} s_{v'}  \le \xi - s_v \indic\{v \in \bar{\cA}(v'')\}    \,\, , \,\, \forall v'' \in V.
$$
In order to compute a sum over $\cN(v)$ recursively, similarly to the case of computing the normalization constant, define $\cQ(s,v,v')$ the set of allowed states in a system where there are $s$ circuits, where a user arriving at beam $v$ does not cause blocking, and where all the beams which are not descendants of $v'$ are empty:
\begin{align*}
\cQ(s,v,v') = \Big\{ \bbn \in \NN^{|V|} &: \sum_{v'' \in \bar{\cA}(v''')} n_{v''} s_{v''}  \le s - s_v \indic\{v \in \bar{\cA}(v''')\}, \\ \forall v''' \in V 
 &\text{ and }  n_{v''} = 0, \forall v'' \not\in \bar{\cD}(v')  \Big\}.
\end{align*}
and we define the partial sums:
$$
	q_{v,v'}(s) = \sum_{\bbn \in \cQ(s,v,v')} \prod_{v'' \in \bar{\cD}(v')} {\rho_{v''}^{n_{v''}} \over n_{v''}!}.
$$
It is noted that since $\cM(\xi,v,1) = \cN \setminus \cN(v)$, we have:
$$
	q_{v,1}(\xi) = \sum_{\bbn \in \cN \setminus \cN(v)} \prod_{v' \in V} {\rho_{v'}^{n_{v'}} \over n_{v'}!},
$$
and
$$
p_v = 1 - { 	q_{v,1}(\xi) \over c(\bbrho)}.
$$ 
so that computing the value of $q_{v,v'}(s)$ for all $v,v' \in V$ and $s \le \xi$ yields the blocking probabilities. It is also noted that if $v'$ and $v$ are not descendants or ancestors of each other, the arrival of a user at beam $v$ does not affect the number of circuits available to the subtree rooted at $v'$, so that we simply have $q_{v,v'}(s) = c_{v'}(s)$, which we already have computed in the first phase of the algorithm. It is also noted that $q_{v,v}(s) = c_v(s-s_v)$. Now, similarly to the previous paragraph, $q_{v,v'}(s)$ obeys the following recursion:
$$
	q_{v,v'}(s) = \begin{cases} \displaystyle \sum_{n_{v'}=0}^{\floor{{s  \over s_{v'}}}} {\rho_{v'}^{n_{v'}} \over n_{v'} !}  \prod_{v'': (v',v'') \in E} & q_{v,v''}(s - n_{v'} s_{v'})   \\
	& \text{ if } v' \in \cA(v) \\
	 c_v(s - s_v)  &\text{ if } v' = v \\
	c_{v'}(s) &\text{ otherwise}
	\end{cases}
$$

{\bf Complexity} It is noted that the required memory for this algorithm to run is $\cO(\xi |V|)$ since, as soon as $p_v$ has been computed at the end of the loop of Phase 2, one can simply remove the values of $q_{v,v'}(s)$ from the memory. It is also noted that the time to compute $c_v(s)$ for $s\le\xi$ and $v \in V$ is $\cO(\xi |E|)$, and, since $G$ is a tree, $|E| = |V|-1$. So the time required for Phase $1$ is $\cO(\xi |V|)$. For Phase 2, for a given value of $v$, $p_v$ is computed by inspecting all the ancestors of $v$, in time $\cO(\xi |\bar{\cA}(v)|) = \cO(\xi h(G))$. Hence the time required for Phase 2 is $\cO( \xi |V| h(G))$. As announced, the algorithm requires time $\cO(\xi |V| h(G))$ and memory $\cO(\xi |V|)$ to compute the blocking probabilities.    	

\section{Numerical experiments}\label{sec:numerical}
Finally we present some numerical experiments to illustrate the flow-level performance of hierarchical beamforming for both elastic and streaming traffic. 
\subsection{Setup}
In order to obtain a realistic setting and take into account the geometry of a cell in a physical network, instead of selecting arbitrary values of $G$, $\bblambda$, $\bbrho$ and $\bbr$, we create a codebook using a similar techniques as in~\cite{tall2015multilevel}, where the main goal is to pack as many beams as possible on each level of the tree, while ensuring that they remain separated with negligible interference. As for the cell geometry, we consider an hexagonal cell, with an antenna height of $30$ meters in a strong Line of Sight environment. We use the standard Nakagami propagation model \cite{charash_1979}, and parameters as done in~\cite{tall2015multilevel}. The output of this process is a graph $G$, as well as the corresponding arrival rate $\bblambda$, loads $\bbrho$ and data rates $\bbr$.

We obtain $G$ a tree with $|V|=10$ beams, and the set of edges is given by $$E = \{(1,2),(1,3),(1,4),(2,7),(2,9),(3,5),(3,6),(4,8),(4,10)\}.$$ It is noted that $G$ is similar to the graph depicted in figure~\ref{fig:map}. The expected flow size is $8$ Mbits, the arrival rates are taken proportional to $$\bblambda = [0.16,0.09,0.12,0.09,0.10,0.10,0.10,0.10,0.09,0.09].$$ The value of $\bblambda$ used here is obtained by assuming that $\lambda_v$ is proportional to the size of the zone covered by $v$. The service rate $r_v$ is given by the harmonic mean of the data rate over the region covered by $v$. The loads are proportional to $$\bbrho = [0.11,0.20,0.31,0.20,0.59,0.59,0.61,0.61,0.18,0.18].$$ In order to avoid scaling problems, in all curves we present the \emph{normalized} flow throughput ${\psi_v \over r_v}$. It is noted that, when the load is very low (respectively very high), the normalized flow throughput should be close to $1$ (respectively $0$).

\subsection{Elastic traffic}
	
\begin{figure}[!h]
\begin{center}
\includegraphics[width=0.85\columnwidth]{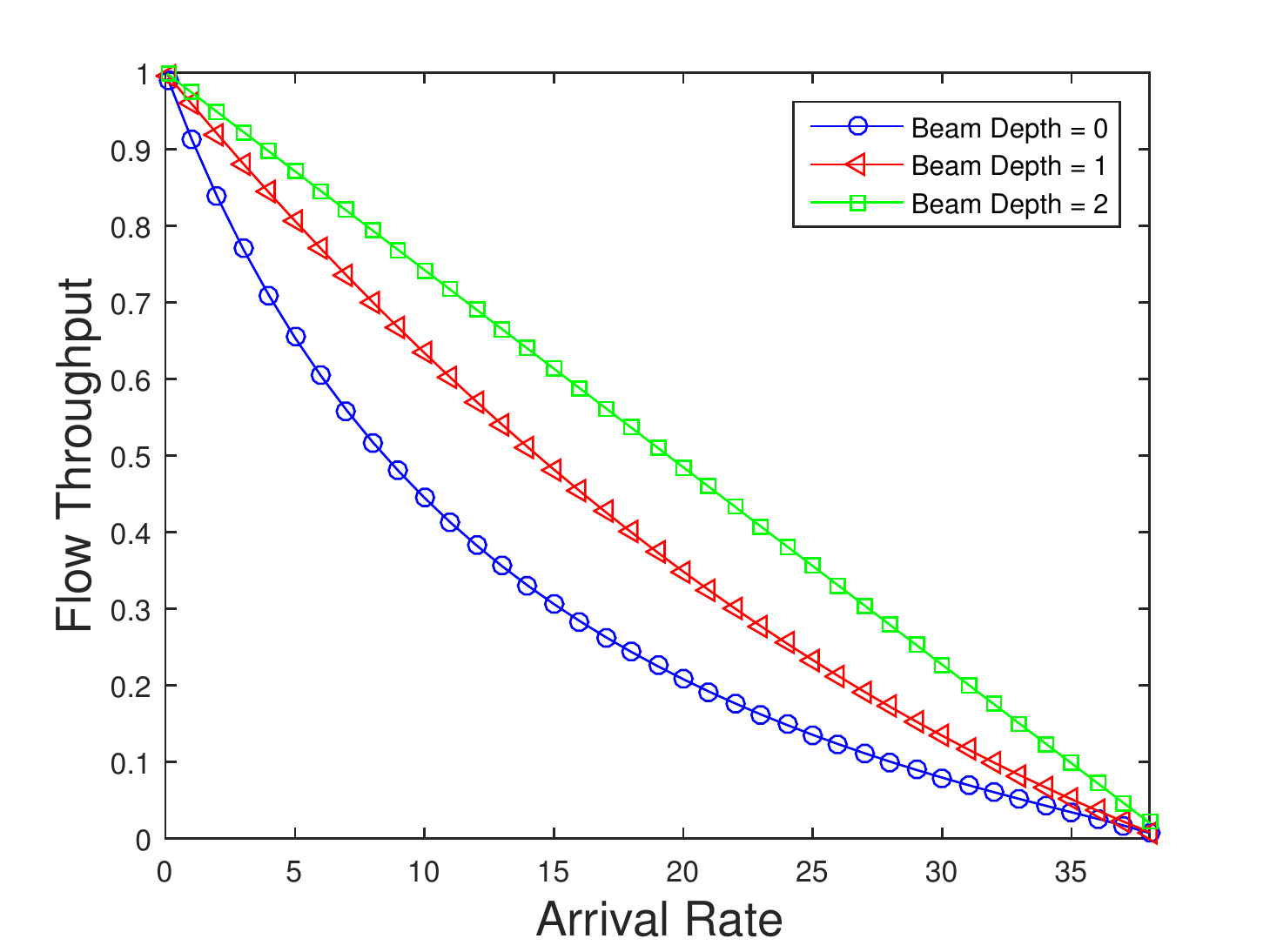}
\caption{Elastic traffic, normalized flow throughput for PF allocation}
\label{fig:PF}
\end{center}
\end{figure}

\begin{figure}[!h]
\begin{center}
\includegraphics[width=0.85\columnwidth]{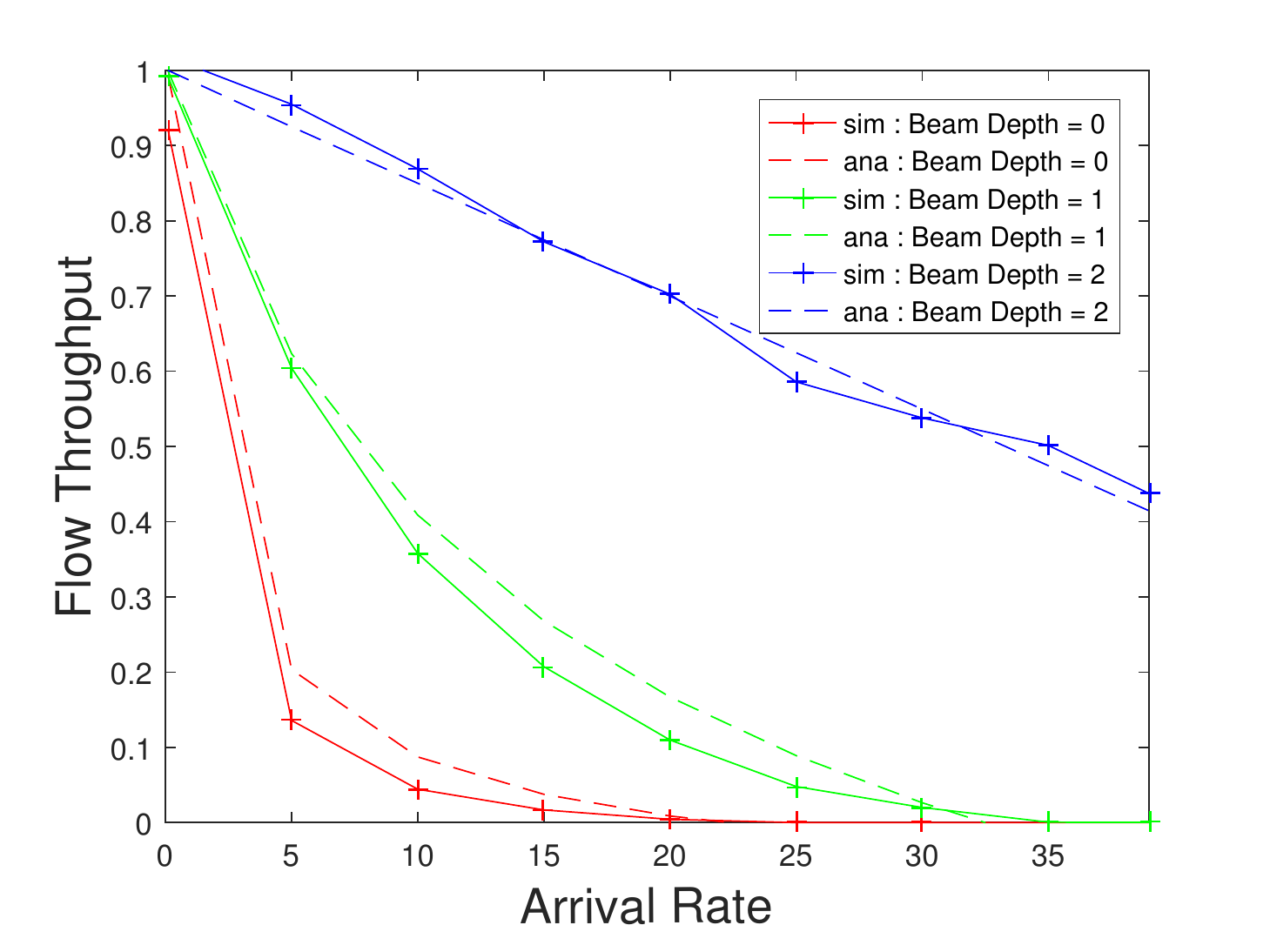}
\caption{Elastic traffic, normalized flow throughput for MT allocation, approximation vs simulations.}
\label{fig:max_throughput_approx}
\end{center}
\end{figure}

	The flow throughput under PF allocation using the tree described above is depicted in Figure~\ref{fig:PF}. Under PF allocation, the flow throughput as a function of the arrival rate is roughly linear for beams of high depth, while it is convex for beams of low depth. The smaller the depth, the higher the curvature. We deduce that PF tends to slightly favor beams with higher depth, and this difference is especially sensible at moderate loads. We can also see in Figure~\ref{fig:PF} that the flow throughput of all beams is strictly positive if $\bbrho$ lies in the stability region. 
The flow throughput under MT allocation using the tree described above is depicted in Figure~\ref{fig:max_throughput_approx}.  In order to assess the accuracy of the approximate formula provided in subsection~\ref{subsec:mt_height2_approx} we show both the approximate analytical expressions as well as the correct value estimated using simulations. In figure~\ref{fig:max_throughput_approx} ''sim'' denotes the values of the flow throughput estimated by simulation, and ''ana'' denotes the approximation given in subsection~\ref{subsec:mt_height2_approx}. It is noticed that the proposed approximation is rather accurate.
On the other hand, under MT allocation, the system is not stable across the whole stability region, and beams of smaller depth get overloaded (so that the flow throughput becomes $0$) when the arrival rate is sufficiently large. This is due to the fact that MT gives absolute priority to beams with high depth. This does not mean that MT is bad per-se:  when flows are static as considered in our model, MT is not throughput optimal, and tends to create overload in beams with small depth. However, in the case where flows are allowed to move during their flow, MT would essentially have an opportunistic behaviour and serve flows only when they can be served by a beam of high depth, at a high data rate. This should dramatically increase the throughput (at the expense of delay, of course). We do not analyze this case here, but it seems rather natural that MT would have such a behaviour.


\subsection{Streaming Traffic}

\begin{figure}[h]
\begin{center}
\includegraphics[width=0.95\columnwidth]{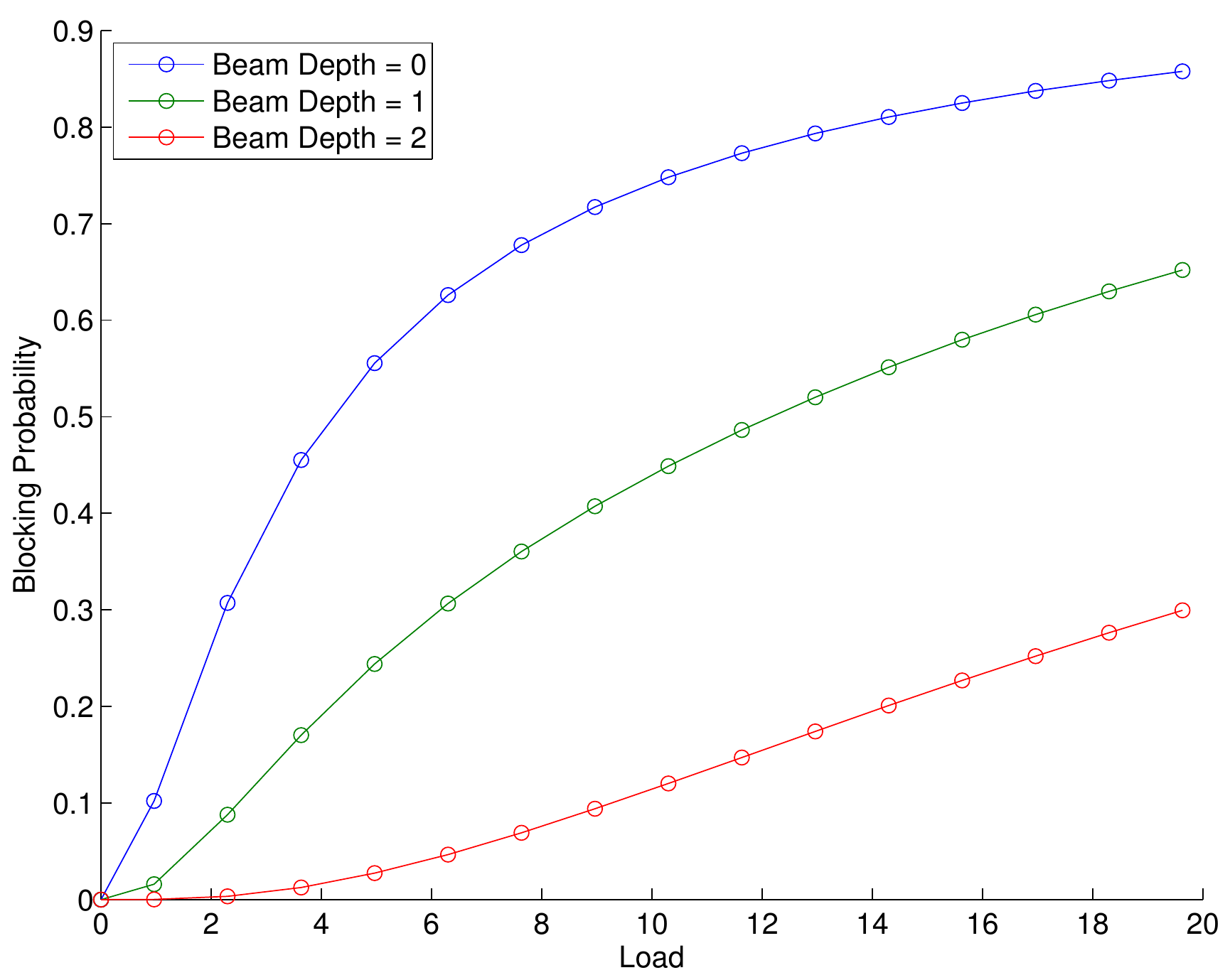}
\caption{Streaming traffic, blocking probability}
\label{fig:streaming}
\end{center}
\end{figure}

We now turn to streaming traffic. We use the same values for the loads $\bbrho$ and the arrival rates $\bblambda$ as in the previous case (up to a scalar multiplicative constant).  On figure~\ref{fig:streaming} we plot the blocking probabilities of the various beams $\bbp(\bbrho)$ with different depth as a function of the system load. As in all Erlang-like models, the blocking rate is almost equal to $0$ for low loads, then rapidly increases once the load reaches a certain threshold. Furthermore, we can see that the system tends to penalize beams with low depth, in particular the root. It would be interesting to design admission control policies which correct that problem, although it does not seem straightforward.
\section{Conclusion}\label{sec:conclusion}
	We have considered hierarchical beamforming in wireless networks, which is an attractive alternative to other beamforming techniques due to the existence of efficient algorithms for flow association and multi-flow scheduling. We have provided computationally efficient algorithms for fair rate allocation including proportional fairness, maximum throughput and max-min fairness, in order to perform resource allocation in real time. We next have proposed closed-form formulas for flow level performance, for both elastic (flow throughput) and streaming traffic (blocking rates). 
\bibliographystyle{acm}
\bibliography{main}
\clearpage
\appendix

\noindent {\bf \LARGE APPENDIX}

\section{Proof of Theorem~\ref{th:alpha_fair_allocation}}

{\bf Stage 1} First consider $\bbgamma \in \conv({\cZ})$ and maximize the objective function with respect to $\bbdelta \in \Delta$. The objective function is separable, so that we obtain $|V|$ independent problems: 
$$
	\text{Maximize } \sum_{k \in \cK(v)} f_\alpha(r_k \gamma_{v} \delta_{k}) \text{   subject to   } \sum_{k \in \cK(v)} \delta_k = 1
$$ 
	From the Karush-Kuhn-Tucker conditions, there exists $\ell_v$ such that the optimal value of $\bbdelta$ satisfies:
	$
		(\gamma_{v} r_k)^{1 - \alpha} \delta_{k}^{-\alpha} = \ell_v 
	$
	for all $k \in \cK(v)$. Using $\sum_{k \in \cK(v)} \delta_{k} = 1$ we get:
$$
	\delta_k^\star = {r_k^{{1 \over \alpha} - 1} \over \sum_{k' \in \cK(v)} r_{k'}^{ {1 \over \alpha} - 1}}
$$
It is noted that $\bbdelta^\star$ is the optimal solution of the original problem since it does not depend on $\bbgamma$. Now set $\bbdelta = \bbdelta^\star$ and it is noted that
\als{
	\sum_{k \in \cK(v)} f(r_{k} \gamma_{v} \delta^\star_{k}) = \phi_v f_\alpha(\gamma_v)  \text{    with     }   
	\phi_v = \Big( \sum_{k \in \cK(v)} r_{k}^{ {1 \over \alpha} - 1} \Big)^{\alpha}.
}

{\bf Stage 2} The original problem reduces to a simpler one:
$$
	\text{Maximize} \sum_{v \in V} \phi_v f_\alpha(\gamma_v) \text{ subject to  } \bbgamma \in \conv(\cZ).
$$
Using Proposition~\ref{prop:conv}, we have that:
$$
	\conv(\cZ) = \left\{  \bbgamma: \gamma_v = \kappa_v \prod_{v' \in \cA(v)}(1-\kappa_{v'}),  \bbkappa \in  [0,1]^{|V|} \right\}
$$
allowing another simplification:
\als{
	\text{Maximize   } & \sum_{v \in V}  \phi_v f_\alpha \Big( \kappa_v \prod_{v' \in \cA(c)}(1-\kappa_{v'})\Big) \\
	\text{subject to   } & \bbkappa \in  [0,1]^{|V|}.
}
 Define $\theta_v$ the optimal value of this optimization problem when only considering $v$ and its descendants, i.e. the value of:
\als{
	\text{Maximize   } & \sum_{v' \in \bar{\cD}(v)} \phi_{v'}  f_\alpha \Big( \kappa_{v'} \prod_{v'' \in \cA(v')}(1-\kappa_{v''})\Big) \\ 
	\text{subject to   } & \bbkappa \in  [0,1]^{|V|}.
}
 Now maximizing with respect to $\kappa_v$:
\als{
	\theta_v = \max_{\kappa_v \in [0,1]} \Big[  \phi_v {\kappa_v^{1-\alpha} \over 1 - \alpha} 
	+ (1-\kappa_v)^{1-\alpha}  \sum_{v':(v,v') \in E} \theta_{v'}  \Big].
}
	Solving the maximization over $\kappa_v$ gives:
$$
	\kappa^\star_v =  \left[ 1 + \left( {1-\alpha \over \phi_v}  \sum_{v':(v,v') \in E} \theta_{v'} \right)^{\frac{1}{\alpha}} \right]^{-1}.
$$
and substituting in the definition of $\theta_v$ we get:
$$
	\theta_v = (\kappa^\star_v)^{1-\alpha} { \phi_v \over  1-\alpha} +  (1-\kappa^\star_v)^{1-\alpha} \sum_{v':(v,v') \in E} \theta_{v'}.
$$
When $v$ is a leaf, we have $\theta_v =  {\phi_v \over 1 - \alpha}$, and the previous relation allows to calculate the value of $\theta_v$ and $\kappa^\star_v$ for all $v \in V$.  Finally, the solution is given by the relation:
$$
\gamma_v^\star = \kappa^\star_v \prod_{v'\in \cA(v)}(1 - \kappa^\star_{v'})  
$$
which concludes the proof.

\section{Proof of Theorem~\ref{th:mt_line}}\label{sec:proof_th_mt_line}

As mentionned above, the system is equivalent to a single M/M/1 queue with FIFO service order and priorities with preemptive resume, where customers of class $v \in V$ represent the users served by beam $V$. Namely, customers of class $v$ receive service if and only if no customers of class $v+1,\dots,|V|$ are present in the system. The result is similar to~\cite{adan2015department}, and we present a proof for completeness. Users of class $|V|$ have the highest priority and are not influenced by other classes so that $N_{|V|}(t)$ follows an M/M/1 process and we have:
$$
\EE( N_{|V|}(t) ) = { \rho_{|V|} \over 1 - \rho_{|V|}}.
$$
 Now consider a class $v < |V|$. We denote by $\EE(W_v)$ the mean amount of time a customer of class $v$ spends in the system, including when receiving service. The waiting time of a customer served by $v$ can be divided in epochs of length $X_1,X_2,...$ as follows. The duration $X_1$ of first epoch equals the amount of work associated with all the customers with the same or higher priority present in the queue upon his arrival, including himself. The length $X_2$ of the second epoch equals the amount of higher priority work arriving during the first epoch (of duration $X_1$). The length of the third epoch equals the amount of higher priority work arriving during the second epoch (of duration $X_2$), etc. This allows to express the waiting time as a sum:
$$ 
\EE(W_v) =\sum_{k\in \NN}\EE(X_k) 
$$
From the PASTA property, the amount of work with equal or higher priority upon arrival of a customer of class $v$ (including himself) is:
$$ \EE(X_1) = {1 \over r_v} + \sum_{v' \ge v} {\EE(N_{v'}(t)) \over r_{v'}}$$
The expected amount of work with higher priority arriving during a duration $X_k$ is given by recursion:
$$\EE(X_{k+1}) = \EE(X_{k}) \left(\sum_{v' > v} \rho_{v'}\right)  = \EE(X_{1}) \left(\sum_{v' > v} \rho_{v'}\right)^{k-1}.
$$
so that, summing:
$$
	\EE(W_v) = {\EE(X_1) \over 1 - \sum_{v' > v} \rho_{v'}}
$$
Replacing $\EE(X_1)$ by its expression:
$$
	\EE(W_v) = {1 \over 1 - \sum_{v' > v} \rho_{v'}} \left( {1 \over r_v} + \sum_{v' \ge v} {\EE(N_{v'}(t)) \over r_{v'}} \right)
$$
From Little's law, $\lambda_v \EE(W_v) = \EE(N_v(t))$ so that:
$$
	{\EE(N_v(t)) \over \lambda_v} = {1 \over 1 - \sum_{v' > v} \rho_{v'}} \left( {1 \over r_v} + \sum_{v' \ge v} {\EE(N_{v'}(t)) \over r_{v'}} \right),
$$
which yields the relation:
$$
	\EE(N_v(t)) = {\rho_v + \lambda_v \sum_{v' > v} {\EE(N_{v'}(t)) \over r_{v'}} \over 1 - \sum_{v' \ge v} \rho_{v'}}.
$$
One may readily check by recursion that:
$$	
	\EE(N_v(t)) = \frac{\rho_v\left(1+\sum_{v' \ge v}\rho_{v'}\left(\frac{r_v}{r_{v'}}-1\right)\right)}{ \left(1-\sum_{v' \ge v }\rho_{v'}\right)\left(1-\sum_{v' > v} \rho_{v'}\right)}   
$$
which is the announced result.

\section{Definitions for trees}
Given a directed tree $G =(V,E)$ we use the following terminology.
\bi
\item   Parent: $v$ is a parent of $v'$ iff $(v,v') \in E$,
\item 	Child: $v$ is a child of $v'$ iff $(v',v) \in E$,
\item   Root: $v$ is the root iff it has no parent.
\item 	Leaf: $v$ is a leaf if it has no children.
\item 	Path: a path from $v$ to $v'$ of length $\ell$ is a set of vertices $(v_0,v_1)$, \dots,$(v_{\ell-1},v_{\ell})$ in $E$ such that $v_0=v$ and $v_{\ell} = v'$.
\item   Depth: the depth of $v$ is the length of the path from the root to $v$.
\item 	Descendant: $v'$ is a descendant of $v$ if there exists a path from $v$ to $v'$
\item 	Ancestor: $v'$ is an ancestor of $v$ if there exists a path from $v'$ to $v$
\item 	Height of the tree: $h(G)$ is the maximal depth of a leaf.
\item   Degree of the tree: $\deg(G)$ is the maximal number of children of a node. 
\item 	Regular tree: a tree is $d$-regular if all nodes have degree $0$ (for leaves) or $d$ (for non-leaves).
\ei